\documentclass[aps, pra, a4paper, showpacs, english, twocolumn, 10pt, nobibnotes]{revtex4-1}
\usepackage{bbm, amsthm, bm,textcomp, nicefrac,geometry,ragged2e}
\usepackage[dvipsnames]{xcolor}
\usepackage{float}
\usepackage[bbgreekl]{mathbbol}
\usepackage{graphicx,epstopdf,color,verbatim,enumitem,ulem}
\usepackage{pbox,array}
\usepackage{mathrsfs}
\usepackage{amssymb}
\usepackage{stackrel}
\usepackage{amsmath}
\makeatletter
\usepackage[caption=false]{subfig}
\usepackage[T1]{fontenc}
\usepackage[caption=false]{subfig}
\usepackage{babel}
\addto\captionsenglish{}

\begin{document}
\title{Genuine quantum networks: superposed tasks and addressing}
\author{J. Miguel-Ramiro$^{1}$, A. Pirker$^{1}$ and W.~D\"ur$^1$}
\affiliation{$^1$ Institut f\"ur Theoretische Physik, Universit\"at Innsbruck, Technikerstra{\ss}e 21a, 6020 Innsbruck, Austria}

\date{\today}

\begin{abstract}

We show how to make quantum networks, both standard and entanglement-based, genuine quantum by providing them with the possibility of handling superposed tasks and superposed addressing. This extension of their functionality relies on a quantum control register, which specifies not only the task of the network, but also the corresponding weights in a coherently superposed fashion. Although adding coherent control to classical tasks, such as sending or measuring ---or not doing so---,  is in general impossible, we introduce protocols that are able to mimick this behavior 
under certain conditions.
We achieve this by always performing the classical task, either on the desired state or a properly chosen dummy state.  We provide several examples, and show that externally controlling quantum superposition of tasks offers new possibilities and advantages over usually considered single functionality. For instance, superpositions of different target state configurations shared among different nodes of the network can be prepared, or quantum information can be sent among a superposition of different paths or to different destinations.

\end{abstract}
\maketitle

\section{Introduction}\label{Intro}

Quantum networks promise to be the backbone of upcoming quantum technologies \cite{Kimble2008,Wehner2018}.
Several tasks have been identified where quantum effects allow one to obtain an advantage over classical approaches, or even make things possible in the first place. Many of these applications are based on the distribution of quantum states to spatially separated parties and by exploiting truly quantum features such as entanglement. This includes security applications such as key distribution \cite{Ekert91,Bennett2014,ShorQKD},
 secret sharing \cite{Markham08,Hillery99} and secret voting \cite{secretvoting1,secretvoting2}, 
 distributed (or cloud) quantum computation \cite{CiracDistributed}, 
 as well as improved sensing or time and frequency standards \cite{Eldredge2018,sekatski2019optimal,Kessler2014}.

There are basically two approaches to such quantum networks: a bottom-up \cite{Meter2011,Wehner2018} and a top-down approach \cite{Pirker2018}. The former one conceptually closely relates to classical networks. In a bottom-up approach, the quantum network completes requests and tasks by sending quantum states through channels, from network device to network device. Even though there are some new elements, such as the generation of {\it quantum} states or the transmission of {\it quantum} information, well-established concepts of classical networks, such as routing or addressing, still appear to be applicable or at least adjustable. The latter approach to quantum networks, i.e. a top-down approach, consists of entanglement-based networks, where devices prepare entanglement beforehand, which is subsequently manipulated in order to complete desired requests. In both cases, stack models \cite{Meter2013a,Pirker2019} that define necessary elements and functionalities have been proposed and analyzed. However, so far the desired functionality of networks is restricted to a specific, classically defined task such as transmitting quantum information to a specific node in the network, or to prepare a certain multipartite entangled quantum state shared among different parties.

In this work we lift the functionality of quantum networks to a genuine quantum level by introducing techniques and procedures which enable network devices to complete tasks and to address other devices in a coherent fashion. Besides, these tasks can in principle be controlled in a quantum way from the outside. This allows for several interesting applications such as the preparation of superpositions of desired target states, possibly shared among different parties. Other applications include the transmission of quantum information to a superposition of different receivers, as well as sending quantum information over a superposition of different paths. Note that this goes beyond multi-path routing as considered e.g. in \cite{Pant2019,Pirandola2019}, where resources are used in a parallel, but not a superposed way. In order to complete tasks in a superposed way, we mimic the behaviour of coherently controlling classical tasks. We remark that adding quantum control to classical tasks, such as performing a measurement, e.g. for state merging or teleportation ---which are part of typical network requests---, is in general impossible, as we argue later. However, we find that one can mimic the behaviour of the system in such a way that the resulting state or network configuration is ``as if'' such a coherently controlled classical operation was performed. This is done by adding quantum control at the level of unitary operations, in such a way that operations act on different desired states or on dummy states, in order to generate the superposition. Crucially, the classical task is always performed. As we show later, one needs to ensure that for the known input state and the dummy state, the probabilities and measurement outcomes are equal and indistinguishable for all involved configurations and states.

We argue that the additional functionality of external quantum control and of handling superpositions of tasks is a desirable and useful feature that offers new possibilities. We illustrate this by providing examples where superpositions of states shared among different parties in the network are generated. This includes e.g. a superposition of three-party GHZ states shared among four parties, where the superposed state can be reduced deterministically to a three-party GHZ state.  In contrast to each of the individual configurations, this superposed state has an additional built-in robustness against losses. Similar observations apply to coherent superpositions of two copies of the four Bell states, which is maximally entangled with respect to all bipartitions. In contrast, each individual state is separable w.r.t. certain bipartitions, and a classical mixture corresponds to the so-called Smolin state \cite{Smolin2001}, which is bound entangled. Further examples include sending of quantum states to a superposition of different locations or in a superposition of different paths, thereby distributing quantum information in a delocalized way within the network, or encoding unknown quantum states within the whole network.

Parallelism in quantum information processing closely relates to adding quantum control to operations, which was investigated in a variety of contexts. For instance, in \cite{Nielsen1997} it was shown that a universal quantum gate array is not feasible, whereas approximate implementations thereof seem to be viable \cite{Vidal2000}. Superposed access to quantum random access memory was investigated in \cite{Giovannetti2008}. Coherently controlling the order of applying unitaries was subject of study in \cite{Araujo2014}, which was experimentally verified in \cite{Procopio2015}, where analyses are performed within the indefinite causal order framework \cite{Oreshkov2012,Chiribella2013}. In addition, the possibility of adding quantum control to unknown operation has been studied in \cite{Friiscontrol,Arajo2014}. The preparation of quantum states in superposition, by applying controlled unitaries, has also received attention. In particular, it has been shown that the so-called {\it quantum adder} for quantum states \cite{AlvarezRodriguez2015,Oszmaniec2016} is, in general, not realizable. However, when partial information of the states is available, a quantum adder turns out to be probabilistically feasible \cite{Oszmaniec2016,adder2}. This has been experimentally investigated in \cite{Hu2016,Li2017}.

In contrast to this former work of adding quantum control to operations, the results we present here differ in two key points. First, we do not require to coherently control the application of all different possible kinds of operations. More precisely, we restrict the set of coherently controlled operations which the network devices apply to be chosen from a {\it finite} set of possible transformations. Second, we do not aim to prepare a superposition of completely unknown states, or to superpose unknown operations. In contrast, we study the {\it distributed} preparation of superpositions of known quantum states by mimicking quantum control of classical tasks. We do not assume the states which shall be brought into superposition to be available a priori, but we coherently control the generation process (unitaries and measurements) which each network device implements onto some network resource. In general the desired process is known, or part of a finite set of possible operations.

The paper is organized as follows. In Sec.~\ref{sec:background} we give an overview of the fundamental concepts and tools we make use of throughout the paper, including characteristics of graph states and quantum networks. We introduce the problem setting in Sec.~\ref{sec:superptasks}, and provide a detailed analysis of the initialization and preparation of the whole network, as well as a general picture of the overall process. All the basic tools and mechanisms to add control to classical tasks are introduced in Sec.~\ref{sec:mimiking}, followed by a detailed example illustrating the procedures required to generate superposition of arbitrary requests within a network (Sec.~\ref{sec:qcrequest}). In Sec.~\ref{sec:addressing}, we show how quantum controlled addressing functionality can also be included within our network approach. Finally, in Sec.~\ref{sec:examples} we provide different examples or scenarios where the generation of superposed states in a coherently controlled way is desirable and beneficial. We summarize and conclude in Sec.~\ref{sec:conclusions}.

\newcommand{\ket}[1]{\left|#1\right\rangle}
\newcommand\ketbra[2]{\left|#1 \right\rangle \left\langle#2\right|}

\newcommand{\px}{\sigma_x}
\newcommand{\py}{\sigma_y}
\newcommand{\pz}{\sigma_z}
\newcommand{\id}{\mathbb{1}}

\section{Background} \label{sec:background}

In this section we provide a brief overview of the relevant background material for this work. In particular, we give a short introduction to Bell-states, GHZ-states and graph states as well as a brief discussion about previous works on quantum networks.

\subsection{Bell-states, GHZ-states and graph states} \label{sec:background:bell}

In the following we make use of  Bell states. These states are two-qubit maximally entangled quantum states. Specifically, the four Bell states are
\begin{align}
\ket{B_{ij}} = (\id \otimes \px^j \pz^i) \ket{B_{00}}, \label{eq:bellbasis}
\end{align}
where $i \in \lbrace 0,1 \rbrace$ is called the phase, and $j \in \lbrace 0,1 \rbrace$ the amplitude bit of the Bell state $\ket{B_{ij}}$ and with $\ket{B_{00}} = (\ket{00} + \ket{11})/\sqrt{2}$. Throughout this paper, we denote the four states $\ket{B_{ij}}$ as $\ket{\Phi_{k}}$ with $k \in \lbrace 0,1,2,3 \rbrace$ and we frequently denote $\ket{\Phi_{0}}\left(\equiv\ket{B_{00}}\right)$ as $\ket{\Phi^{+}}$ for simplicity.

Such maximally entangled states are a valuable resource for different applications in a distributed setting, including e.g. super-dense coding \cite{Bennett92} and quantum teleportation \cite{Bennett93}. We briefly recall the steps comprising the teleportation protocol, since we will require them later in this work. In quantum teleportation, two communication partners, Alice and Bob, share a perfect Bell pair in the state $\ket{\Phi^+}$. If now Alice wants to transmit an unknown single-qubit state $\ket{\varphi}$ to Bob, she performs a Bell measurement between the qubit to be transmitted and her half of the Bell pair, and sends the outcomes of the measurement classically to Bob. This in turn enables Bob to restore the state $\ket{\varphi}$ on his qubit by performing a local Pauli correction operation that depends on the measurement outcome.

GHZ states are the natural extension of Bell states to more than two parties. We define a $n$-qubit GHZ state as
\begin{align}
	\ket{\mathrm{GHZ}_n} = \frac{1}{\sqrt{2}}\left( \ket{0}^{\otimes n} + \ket{1}^{\otimes n} \right).
\end{align}
GHZ states are useful for applications such as clock-synchronization \cite{Komar14}, distributed sensing \cite{Eldredge2018} and quantum key agreement \cite{XuQka}.

Graph states are $n$-qubit quantum states which exhibit correlations corresponding to classical graphs \cite{He06}. 
Generally speaking, graph states are so-called stabilizer states, i.e. states which are stabilized by elements of the Pauli group. Precisely, given a classical graph $G=(V,E)$, where $V$ denotes the set of vertices and $E$ the set of edges, the graph state $\ket{G}$ is defined as the unique $+1$ eigenstate of the set of operators
\begin{align}
K_{a} = \px^{(a)} \prod_{(a,b) \in E} \pz^{(b)}, \label{def:graph}
\end{align}
for all $a \in V$, where the superscript indicates on which qubit the Pauli operator is acting on. In other words, $K_{a} \ket{G} = \ket{G}$ for all $a \in V$ and $K_{a}$ as defined in Eq. (\ref{def:graph}). One easily verifies that the state $\ket{G}$ can also be explicitly written as
\begin{align}
\ket{G} = \frac{1}{\sqrt{2}}\left(\ket{0}_{a} \ket{G / a} + \ket{1}_{a} \prod\pz^{N(a)} \ket{G / a} \right), \label{obs:vertexdecomp}
\end{align}
for any vertex $a \in V$, where $N(a)$ refers to the neighbourhood of vertex $a$. This decomposition turns out to be useful e.g. when merging two graph states.

\subsection{Transformations of entangled states}
We make use of different quantum operations acting on entangled states that we recall here.

\subsubsection{Bell-measurement} \label{sec:bellmeas}
A Bell-measurement is a joint measurement between two qubits, that can be part of some larger entangled state, such that the joint state of the qubits is projected into one of the elements of the Bell basis (Eq. \ref{eq:bellbasis}). We consider Bell-measurements between different states. First, we consider a Bell-measurement between two GHZ-states of arbitrary size, say states $\ket{\mathrm{GHZ}_m}$ and $\ket{\mathrm{GHZ}_n}$. The state after the measurement is, up to local correction operations, given by $\ket{\mathrm{GHZ}_{n+m-2}}$. On the other hand, we make use of a Bell-measurement  between an arbitrary single-qubit state, e.g. $\ket{\varphi} = \alpha \ket{0} + \beta \ket{1}$, and a GHZ-state of size $n+1$, i.e. $\ket{\mathrm{GHZ}_{n+1}}$. The state after the measurement reads as
\begin{align}
\alpha \ket{0}^{\otimes n} + \beta \ket{1}^{\otimes n},
\end{align}
up to local corrections of the form $\left\lbrace\id, \pz^{\otimes n}, \px^{\otimes n}, \px^{\otimes n}\pz^{\otimes n}\right\rbrace$ for each measurement outcome $\ket{\Phi_i}$.

An extension to qudit systems is straightforward, and results in a state of the form
\begin{align}
\sum\limits^{d-1}_{i=0} \alpha_{i} \ket{i}^{\otimes n},\label{GHZn}
\end{align}
where one uses a $d$-level GHZ state, i.e. a state of the form Eq. (\ref{GHZn}) with $\alpha_i=\sqrt{d}$ and $n+1$ systems as input.

We also observe that a GHZ-state is local unitary (LU) equivalent to a graph state. Specifically, we can transform a GHZ state of size $n$ into a graph state by the following local transformation. A GHZ-state of size $n$ is stabilized by operators of the form $\pz^{(1)} \otimes \pz^{(i)}$ for $2 \leq i \leq n$ and $\px^{\otimes n}$. Now suppose that we apply a Hadamard rotation to all qubits except the first. Then, because $H \pz H = \px$, the stabilizers transform to $\pz^{(1)} \otimes \px^{(i)}$ for $2 \leq i \leq n$ and $\px \otimes \pz^{\otimes (n-1)}$, which corresponds to the stabilizers of the star graph state.

\subsubsection{Cutting of graph states}
Graph states show a simple behavior under Pauli measurement of single qubits, which can be described by graphical rules on the corresponding graph. Consider a graph state of the form Eq.~(\ref{obs:vertexdecomp}). A measurement with respect to the Pauli $\pz$ operator on qubit $a$ has the effect that qubit $a$ is cut from the rest of the graph state and the resulting state is $\ket{G / a}$, up to  $\prod\pz^{N(a)}$ corrections. Other Pauli measurements lead to additional changes of the resulting graph state, see  \cite{He06} for details.

\subsubsection{Merging of graph states} \label{sec:backmerg}
Consider two graph states, $\ket{G_1}$ and $\ket{G_2}$, of the form Eq.~(\ref{obs:vertexdecomp}). We want to merge the vertices $a_1 \in V_1$ and $a_2 \in V_2$ into a single vertex $\tilde{a_{1}}$. For that purpose we measure $a_1$ and $a_2$ with respect to the operators $P_0 = \ketbra{0}{00} + \ketbra{1}{11}$ and $P_1 = \ketbra{0}{01} + \ketbra{1}{10}$. Assuming we find the measurement outcome $0$  w.r.t. $\lbrace P_0, P_1 \rbrace$, the resulting state reads
{\small{}
\begin{multline}
\left|G_{1}\right\rangle \left|G_{2}\right\rangle \overset{\left\{ P_{0},P_{1}\right\} }{\longrightarrow}\\
\frac{1}{\sqrt{2}}\left(\left|0\right\rangle _{\tilde{a_{1}}}\left|G_{1}\cup G_{2}/\tilde{a_{1}}\right\rangle +\left|1\right\rangle _{\tilde{a_{1}}}\prod\sigma_{z}^{N_{a_{1}}\cup N_{a_{2}}}\left|G_{1}\cup G_{2}/\tilde{a_{1}}\right\rangle \right),\label{eq:mergedgraphs}
\end{multline}}where the state is renormalized. In case the outcome  $1$ is found in the measurement, one can restore the state of Eq. (\ref{eq:mergedgraphs}) by applying a correction operation of the form $\prod\pz^{N(a_2)}$. That is, the resulting state corresponds to a graph state $\left|G_{1}\cup G_{2}\right\rangle$ where the two vertices $a_1$ and $a_2$ are merged into one vertex denoted as $\tilde{a_{1}}$.

\subsection{Quantum networks \label{sec:background:networks}}

The construction of large-scale quantum networks involves several obstacles that need to be overcome. For instance, sending quantum states directly over unconditionally long distances is not possible due to the No-Cloning theorem \cite{Wootters82}. This obstacles are addressed by so-called quantum repeaters \cite{Br98,DurRepeater}, which enable for long-distance quantum communication. Different approaches for building quantum repeaters exist, such as by directly utilizing channels and using quantum error correction \cite{Zw14,Muralidharan2014}, or by exploiting bipartite  \cite{Zwerger18,Pirandola2017},  and multipartite entanglement \cite{Wallnofer16_2D,Wallnofer2019}. Quantum networks utilize quantum repeaters to generate entanglement over arbitrary distances. Quantum networks are also constructed by different approaches, e.g. by using bipartite entanglement (also referred to as quantum repeater networks \cite{Pirandola16,Meter2013a,Meter2013b,Meignant2019}), and by using multipartite entanglement \cite{Epping2016a,Hahn2019,Pirker2018,Pirker2019}. In addition, noise and imperfections in transmission channels and network devices have to be tackled. This is the subject of study in fault-tolerant quantum computation \cite{Chamberland18,Chao18}, quantum error correction and entanglement distillation protocols \cite{Deutsch96,Bennett96}. Finally, also the organization, management, operation and design of quantum networks poses a significant challenge \cite{Pirker2018,Meter2013a,Pirker2019,Meter2011}.

Two different approaches how to organize and build quantum networks exist: bottom-up \cite{Meter2011}  
  or top-down \cite{Pirker2018,Pirker2019}. In a bottom-up approach quantum networks combine the resources of the network, e.g. quantum channels or entanglement, depending on the task, in an appropriate manner. For example, suppose that three clients of a quantum network request to share a three qubit GHZ-state. In a bottom-up approach, the quantum network devices need to route and make use of the local resources to fulfill the request \cite{Meter2013b,Pirandola16,Hahn2019,Gyongyosi2017,Gyongyosi18}.  In contrast, quantum networks using a top-down approach first prepare a universal resource which the devices use later to complete all required tasks. For that purpose, quantum networks mainly use multipartite entangled quantum states. If now clients issue a task to the network, the devices manipulate these states according to the task. Top-down quantum networks minimize the waiting times for clients and result in states with higher fidelity (due to less merging). However the network devices need to prepare the universal resource beforehand, and store them until the task should be performed.

In principle, quantum networks can complete different tasks. In this work we focus on two main tasks: the transmission of quantum information between two (or more) distant communication partners, and the generation of multipartite entangled states shared among different clients \cite{Pirker2018,Pirker2019,Meter2013a,Spee2013}.

Some of the concrete (sub)tasks we investigate in this work include:

\begin{enumerate}[label=(\roman*)]
\item Sending of quantum information by means of quantum teleportation \cite{Bennett93}.
\item Sending of qubits via quantum channels.
\item Sending of quantum information through certain paths of a network.
\item Distribution of quantum information among network devices.
\item Preparation of certain multipartite entangled states between arbitrary network devices under request, including state manipulation (e.g. cutting and merging of graph states).
\item Addressing of network devices.

\end{enumerate}
There exist works which study these tasks in detail. For instance, in \cite{Meter2013b} it was studied how to determine paths in quantum repeater networks. In contrast, Refs. \cite{Epping2016a,Pirker2018} study how to generate graph states in quantum networks. However, all of these works have in common that they investigate a single task. In this work we provide  functionalities that empower a quantum network such that it is able to also complete these tasks in a \textit{coherent superposition}. As we will show, there are several examples in which it is useful and beneficial to complete tasks in a coherent superposition, compared to completing them individually or considering the corresponding classical mixture of tasks. 

\newcommand\braket[2]{\langle#1|#2\rangle}  

\def\dm#1{\left|#1 \right\rangle \left\langle #1 \right|}

\section{Superposition of tasks in quantum networks}\label{sec:superptasks}

In the following we outline the  problem setting we consider in this work, as well as the general idea about how we tackle the problem.

For that purpose we consider a quantum network that comprises $n$ quantum network devices. The network devices connect in an arbitrary manner, either by some entangled resource state or via quantum channels. We illustrate our approach for entanglement-based quantum networks throughout the paper, as this case is conceptually simpler. We show later how to extend it to other situations and settings.

We summarize the entanglement resource of the quantum network as the state $\ket{\psi}_{res}$. Additionally, we denote as $\ket{\psi}_{aux}$ the global state of the auxiliary qubits belonging to the network devices. These auxiliary qubits are systems locally prepared by each device in a suitably way, and the number of auxiliary qubits stored depends on each task and scenario. The goal of the quantum network is to enable for the coherent completion of different tasks, such as those mentioned above in Sec. \ref{sec:background:networks}. 
We first consider tasks in a limited sense, where we deal with the preparation of superpositions of quantum states. We discuss later if and how this can be extended to more general settings. By this we mean also the superpositions of "applications", e.g. superpositions of sending and not sending, sending among different paths, encoding of information into superposition of different codes, or performing a BB84 protocol in a superposed way.

Suppose that the request for the quantum network is to prepare a superposition of $m$ different tasks represented by quantum states, i.e. the states $\ket{\psi_{1}}, \ldots, \ket{\psi_{m}}$, with weights $\alpha_{1}, \ldots, \alpha_{m} \in \mathbb{C}$. Precisely, the state which shall be prepared by the  network devices (the ones actively involved in the realization of the particular tasks), referred to as target state, reads as
\begin{align}
\ket{\psi_{T_{1}}} = \sum\limits^{m-1}_{i=0} \alpha_{i} \ket{i}_{c} \ket{\psi_{i}},
\label{eq:targetstate}
\end{align}
where each state $\ket{\psi_{i}}$ defines the completion of each particular task and can involve resource as well as auxiliary systems. In some cases, the external quantum control  (sub-index $c$) can be deterministically detached, leading to a target state of the form
\begin{align}
\ket{\psi_{T_{2}}} = \sum\limits^{m-1}_{i=0} \alpha_{i} \ket{\psi_{i}}. \label{eq:targetstate2}
\end{align}

\begin{figure}
\includegraphics[width=\columnwidth]{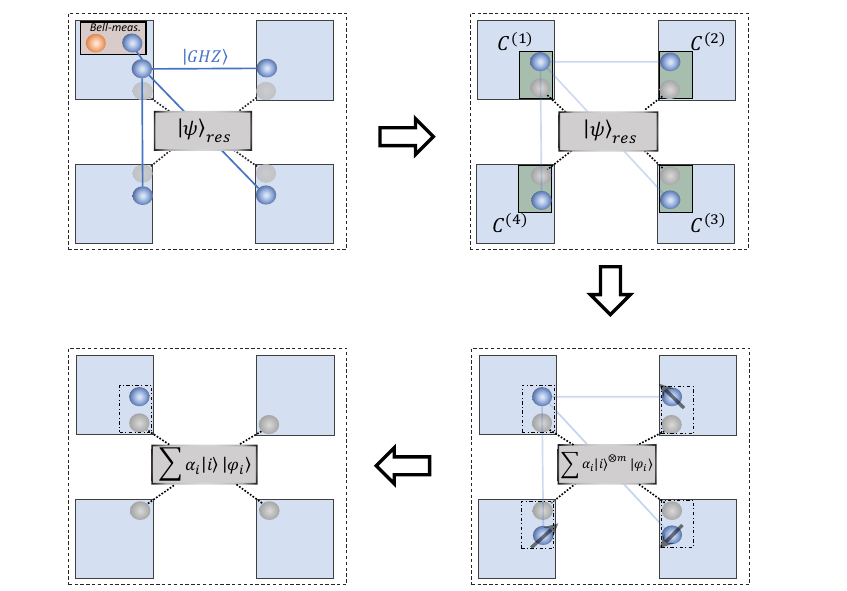}
\caption[h!]{\label{fig:setting} Schematic illustration of the overall process for a four-device quantum network. Upper-left device acts as initiator.  Each grey vertex represents all the resource and auxiliary qubits that each device owns. First (upper-left), the initiator device prepares the weight state of Eq. (\ref{eq:weightstate}), which corresponds to the orange vertex. Further, we require that the four quantum network devices share a five qudit GHZ-state, see Eq. (\ref{eq:prerequeststate}), which comprises the blue vertices. After the Bell-measurement, the four quantum network devices share the state $\ket{\psi_{R}}$ of Eq. (\ref{eq:requeststate}). In second place (upper-right), each device applies the corresponding controlled unitaries, using the request state $\ket{\psi_{R}}$ and their auxiliary qubits $\ket{\psi}_{aux}$, to the resource state of the network $\ket{\psi}_{res}$ in a coherent way. The maps $C^{(i)}$ are defined as $C^{(i)} = \prod^{m-1}_{k=0} C^{(i)}_{k}$, i.e. the product of all controlled unitaries for one particular network device. In this way, the desired coherent superposition is generated (bottom-right). Finally, all control registers, except the initiator one, are measured and the desired superposed state (up to corrections) of Eq.~(\ref{eq:aftercontrolledunitariesphi}) is generated (bottom-left).}
\end{figure}

In order to prepare the states of Eq. (\ref{eq:targetstate}) and Eq. (\ref{eq:targetstate2}), we propose the following procedure. The tasks the network should perform are specified by a quantum state of a single qudit that one of the parties prepares or receives from outside,
\begin{align}
\sum\limits^{m-1}_{i=0} \alpha_{i} \ket{i}, \label{eq:weightstate}
\end{align}
which we also refer to as weight state. The coefficients $\alpha_{i} \in \mathbb{C}$ of this state specify the weights  of the superposed target tasks $\ket{\psi_{i}}$,  and is supplemented by additional classical information on the operations to be performed. Together with a previously shared $n+1$ qudit GHZ-state
\begin{align}
\frac{1}{\sqrt{m}} \sum\limits^{m-1}_{i=0} \ket{i}^{\otimes (n+1)}, \label{eq:prerequeststate}
\end{align}
the parties can prepare the required control state via Bell measurement by the initiator device (Sec. \ref{sec:bellmeas}):
\begin{align}
\ket{\psi_{R}} = \sum\limits^{m-1}_{i=0} \alpha_{i} \ket{i}^{\otimes n}_{c}, \label{eq:requeststate}
\end{align}
We refer to this resulting state as request state.  Quantum control of further operations is determined by this request state.  The scheme is depicted in Fig. \ref{fig:setting}. Note that the size of the  state Eq. (\ref{eq:requeststate})
depends on the number of devices of the network that take part in this process, and its dimensionality
depends on the number of constituents of the final superposition.

An important observation in Eq. (\ref{eq:requeststate}) is that each quantum network device stores exactly one qudit. This enables each network device to apply controlled unitaries on the resource and auxiliary states $\ket{\psi}_{res}\ket{\psi}_{aux}$ of the quantum network. We imitate the behaviour of controlled-tasks by suitably adding control at the level of unitary operations. More specifically, the request state of Eq. (\ref{eq:global1}) enables the quantum network device $j$  to apply controlled unitaries $U_{i}$ for $0 \leq i \leq m-1$, i.e. operations of the form
\begin{align}
C^{(j)}_{i} = (\id - \dm{i}) \otimes \id^{(j)} + \dm{i} \otimes U^{(j)}_{i}. \label{eq:controlledunitary}
\end{align}
Each device is provided beforehand, together with Eq. (\ref{eq:weightstate}), with a classical description of what unitary they have to apply for each state of the control register.

Unitaries $U^{(j)}_{i}$ for $0 \leq j \leq n-1$ are coherently applied by all quantum network devices, i.e. the application of the controlled unitaries of Eq. (\ref{eq:controlledunitary}) for $0 \leq i \leq m-1$ and all $0 \leq j \leq n-1$ to the resource and auxiliary state $\ket{\psi}_{res}\ket{\psi}_{aux}$ results in the state
\begin{align}
\ket{\psi_{C}} &= \prod\limits^{m-1}_{k=0} \bigotimes\limits^{n}_{j=1} C^{(j)}_{k} \sum\limits^{m-1}_{i=0} \alpha_{i} \ket{i}^{\otimes n}_{c} \ket{\psi}_{res}\ket{\psi}_{aux} \notag \\
&= \sum\limits^{m-1}_{i=0} \alpha_{i} \ket{i}^{\otimes n}_{c} U^{(1)}_{i} \otimes \ldots \otimes U^{(n)}_{i} \ket{\psi}_{res}\ket{\psi}_{aux}. \label{eq:controlledu1}
\end{align}
Therefore, the request state $\ket{\psi_{R}}$ of Eq. (\ref{eq:requeststate}) enables the network devices to apply unitaries in a coherently, controlled and synchronized manner (see also Fig. \ref{fig:setting}). These unitary operations are applied on both, the resource and the auxiliary qubits, which are adequately prepared by each device.

In a next step, all quantum network devices except the initiator device measure their qudits w.r.t. the generalized Pauli $\sigma_{x}$ observable, such that $\left|k\right\rangle _{c}^{\otimes n}\rightarrow\left|k\right\rangle _{c}$, up to phases. In this way, the initiator device becomes the only one still holding the control system of the resulting state. One straightforwardly verifies that we can always transform the resulting state to
\begin{align}
\sum\limits^{m-1}_{i=0} \alpha_{i} \ket{i}_{c} U^{(1)}_{i} \otimes \ldots \otimes U^{(n)}_{i} \ket{\psi}_{res}\ket{\psi}_{aux}, \label{eq:aftercontrolledunitaries}
\end{align}
by applying local corrections consisting of some phases that can be corrected by simply acting on the remaining control register. By defining $\ket{\varphi_{i}} = U^{(1)}_{i} \otimes \ldots \otimes U^{(n)}_{i} \ket{\psi}\ket{\psi}_{aux},$ we can rewrite Eq. (\ref{eq:aftercontrolledunitaries}) to
\begin{align}
\sum\limits^{m-1}_{i=0} \alpha_{i} \ket{i}_{c} \ket{\varphi_{i}}, \label{eq:aftercontrolledunitariesphi}
\end{align}
where the states $\ket{\varphi_{i}}$ involve the resource and auxiliary qubits in a non-trivial way.

The suitable application of the unitaries in a controlled way intends to imitate the behaviour of certain tasks in a coherent way. As we show later, in order to accomplish this and generate the target state $\ket{\psi_{T_{1}}}$ of Eq. (\ref{eq:targetstate}), such that states $\ket{\varphi_{i}}$ relate to the states $\ket{\psi_{i}}$, each particular task has to always be implemented. The implementation of the task usually involves measurements, such as a Bell-measurement or a merging measurement. The crucial point is if these measurements are applied on desired or on dummy states. In this way, adding control at the level of unitaries allows us to effectively add quantum control at the level of tasks. We remark that the unitaries are known, and correspond to SWAP operations.

In case that one wants to get rid of the control qubit and end up with states of the form Eq. (\ref{eq:targetstate2}), we observe that, in general, this is not possible. 
A measurement of the generalized Pauli $\sigma_{x}$ observable on the qudit of the initiator device may lead to a change of the weights $\alpha_{i}$ of the superposition in Eq. (\ref{eq:targetstate2}), therefore jeopardizing the coherence. However, in case that the states $\ket{\psi_{1}}, \ldots, \ket{\psi_{m}}$ are mutually orthogonal, one obtains with probability $1/m$ the state $\sum^{m-1}_{i=0} \alpha_{i} e^{\chi_{i}} \ket{\psi_{i}}$ via a generalized Pauli $\sigma_{x}$ measurement, up to unwanted phases $\chi_{i}$. In some cases these phases can be corrected using local operations on the remaining systems, although this is not always possible. Hence, orthogonality of the final constituents turns out to be a crucial property. In Sec. \ref{sec:extralevel}, we propose a procedure that in some relevant cases allows us to guarantee this orthogonality and correct unwanted phases, therefore being able to get rid of the control register deterministically. In all other cases, additional entanglement is required to resolve this issue.

\subsection{Fully quantum description} \label{sec:fullyq}
So far, we have assumed that the description of the desired target states and the required actions is given classically for all branches. This is typically the case for all single-task requests in networks, and hence it is also natural to assume this for the superposition of tasks. 

However, we point out that this description can be made fully quantum. In order to achieve this, a program register is attached and sent to the devices, which encodes the information
of the actions to be performed by each device, depending on the request state. Thus, the global state can be defined as
\begin{equation}
\ket{\psi_{C}}=\left(\sum_{i=0}^{m-1}\alpha_{i}\left|i\right\rangle _{c}^{\otimes n}|\phi_{U}^{(i)}\rangle ^{\otimes n}\right)\left|\psi\right\rangle_{res}\ket{\psi}_{aux}.\label{eq:global1}
\end{equation}
This program register has to be attached together with the request qubits that are distributed among the network devices, by modifying the GHZ state construction of Eqs. (\ref{eq:weightstate})-(\ref{eq:requeststate}). The first register of Eq. (\ref{eq:global1}) is the
request register of Eq.~(\ref{eq:requeststate}), a bit-string data register which defines  the operations applied in each case.
The second register is the  aforementioned program register.  The program register encodes the information of all the unitary operations needed, and is implemented in each device $j$ by a
programmable quantum array gate, in analogy to \cite{Nielsen1997,Vidal2000}. It has the following effect:
\begin{equation}
G^{(j)}\left[\left|\psi\right\rangle\otimes\left|\phi_{U}\right\rangle \right]=\left(U_{k}^{j}\left|\psi\right\rangle\right)\otimes\left|R_{U_{k}^{j}}\right\rangle, \label{eq:gprog}
\end{equation}
where $\left|R_{U}\right\rangle$ is some residual state. Essentially, for any input control state, it invokes the operation $\prod_{k}\left|k\right\rangle_{c} \left\langle k\right|\otimes U_{k}^{(j)}$, where $U_{k}^{j}$ acts locally on the resource and auxiliary qubits of the $j$ network device. Note that, following \cite{Nielsen1997,Vidal2000},
a deterministic programmable array is realizable when considering a finite number of tasks, e.g. the generation of superpositions of graph states, since in this case we deal with a finite number of unitary transformations to be encoded and invoked by the program register. When demanding full functionality, i.e. an infinite number of possible tasks asuch as the generation of all possible target states, the restrictions of programable gate arrays apply \cite{Nielsen1997,Vidal2000}. Observe also that, even for a finite number of tasks, the residual states have to be taken into account during the rest of the process and, in principle, cannot be detached deterministically. 

\section{Mimicking quantum controlled classical tasks} \label{sec:mimiking}

In this section we show indications that adding quantum control to classical tasks
in a way that is desirable for our purpose is in general impossible. We introduce an approach
based on controlled unitary operations that allows us to overcome this problem
and imitate the effect of different controlled tasks, including controlled measurements on partially known states or controlled sending of information, in
a coherent way. We show applications of our approach that allow one
to effectively add quantum control to different classical processes.

\subsection{Coherent controlled measurements on arbitrary pure states} \label{sec:cmeasur}
The feasibility of adding quantum control to quantum measurements has not been explored previously. Given the fact that adding control to unknown unitaries is in general impossible \cite{Arajo2014,Chiribella2013,Friiscontrol}, one can expect similar no-go results for adding control to measurements. In addition, a measurement is by definition an incoherent process, which poses additional challenges when attempting to add control in a coherent way.

The first challenge is already a proper definition of the desired functionality, i.e. how one formally defines a controlled-measurement operation. A formal discussion about it goes beyond the purpose of this paper. We restrict ourselves to one particular desired effect of a transformation that can be interpreted as a certain kind of controlled projective measurement acting on pure states. 
Several indications show that the transformation we require is not a valid quantum operation in general (see Appendix A). However, we also show that we can actually mimick the desired behavior on pure states, which is sufficient for our purpose. In the following we consider performing known measurements on unknown quantum states, and adding control to this process. We will later restrict to performing known measurements on partially known quantum states.

The desired effect of the transformation is to obtain a coherent superposition of a state being measured or not, depending on the state of an additional quantum control register. In particular, if the control register is $|0\rangle$, the input state should remain the same. If the control register is $|1\rangle$, a particular, pre-defined measurement should be performed on the input state. A measurement is however a stochastic process, where with certain probability one out of several outcomes is obtained. What we actually demand is that for each of the possible outcomes of the measurement (which we also denote as branches), we obtain a coherent superposition of the unperturbed state, and the properly renormalized state after obtaining this particular measurement outcome, in such a way that the weights in the superposition are the same for all branches. In addition, each of the branches should happen with the probability $p_k$, that corresponds to the measurement outcome $k$.

Therefore, consider two qubit registers. The first one, the control register, is given by the state
\begin{equation}
\left|\varphi\right\rangle _{c}=\alpha_{0}\left|0\right\rangle +\alpha_{1}\left|1\right\rangle,
\end{equation}
where coefficients $\alpha_{0}$ and $\alpha_{1}$ define the weights of the desired superposition. The measurement is performed in a controlled way on some pure target state $|\psi\rangle$. We consider a POVM $\{A_k\}$ with $\sum_k A_k^\dagger A_k =1$. When obtaining an outcome $k$, the state after the measurement is given by $|\psi_k\rangle=A_k |\psi\rangle/\sqrt{p_k}$, which occurs with probability $p_k=\langle \psi| A_k^\dagger A_k |\psi\rangle$.
The desired effect for an arbitrary projective measurement $M$ is thus
\begin{equation}
|\varphi\rangle_c|\psi\rangle_t \rightarrow \{p_k, \alpha_0|0\rangle_c|\psi\rangle_t + \alpha_1|1\rangle|\psi_k\rangle\}.\label{meas}
\end{equation}

Notice that we have left out an additional register for the state of the measurement apparatus, as one would usually include in a formal description of the measurement process. In a standard description, including the state of the measurement apparatus, the target state for a particular result of the measurement would read as $\alpha_{0}\left|0\right\rangle _{c}\left|\psi\right\rangle _{t}\left|0\right\rangle _{m}+\alpha_{\text{1}}\left|1\right\rangle _{c}|\psi_k\rangle_t|k\rangle_m$. The states $|k\rangle_m$ of the measurement apparatus indicate different measurement outcomes, where $|0\rangle_m$ corresponds to the case where no measurement is performed. All states of the measurement apparatus are mutually orthogonal and classical (and can hence be copied). This implies that one would actually obtain an {\it incoherent} mixture of the unmeasured state and the different measurement branch. One may circumvent this problem ---as we do later--- by actually always performing a measurement, either on a dummy state to preserve the input state $|\psi\rangle$, or on $|\psi\rangle$. In this case the resulting state is
$\sum_k \sqrt{p_k}(\alpha_{0}\left|0\right\rangle _{c}\left|\psi\right\rangle _{t}+\alpha_{\text{1}}\left|1\right\rangle _{c}|\psi_k\rangle)|k\rangle_m$, i.e. the state of the measurement register factors out.

We give some further details in Appendix A. In general, the transformation of Eq. (\ref{meas}) is however non-linear, and can hence not be realized by a quantum mechanical process. There is also an inconsistency for mixed input states. If one takes this desired behavior ---that is only defined for pure states--- to derive the action on mixed states, this action is actually not well defined. When we assume linearity (i.e. the existence of a quantum mechanical process that can realize the desired behavior in general) and consider two equivalent descriptions of a mixed state using different basis states, one obtains different predictions for the target state. We take this as indications that adding control even to known measurements is in general impossible. We leave a formal description and discussion to further work.

\subsection{Controlled measurements on known states}
\begin{figure}[h!]
\includegraphics[width=\columnwidth]{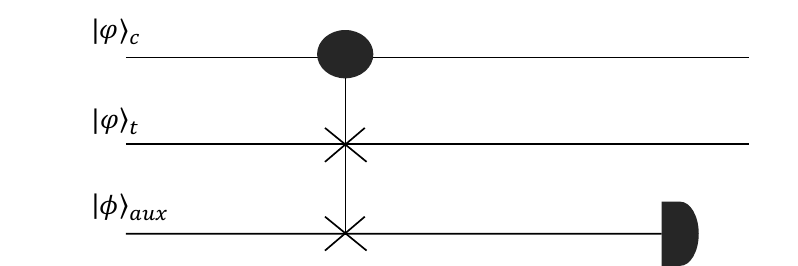}
\caption[h!]{\label{fig:cmeas} Controlled measurement is performed by a controlled-swap operation, also known as Fredkin gate \cite{Fredkin82,Patel_2016}, followed by the measurement of the auxiliary qubit. The auxiliary state has to be suitably prepared depending on the measurement basis and the target state, in order to guarantee that the weights of the final superposition do not change.}
\end{figure}

Although the transformation of the previous section seems in general not physically
realizable, we show how one can effectively reproduce its effect in
a suitable way. To this aim we consider a two-outcome projective measurement with a qubit control register for simplicity. However an extension to general measurements is straightforward.
Consider three qubit registers, a control, a
target and an auxiliary register. The process now consists of the following steps. First, we apply a controlled swap operation, also known as Fredkin gate \cite{Fredkin82,Patel_2016}, acting on the target and the auxiliary qubit, which is controlled by the control qubit (see Fig. \ref{fig:cmeas}). After that, we measure the auxiliary qubit. In order to induce the coherent superposition from the control qubit, we need to choose the auxiliary qubit accordingly. Note that we denote here the input state to be measured as target state.

The initial global state is given by $\left|\Psi\right\rangle =\left|\varphi\right\rangle _{c}\otimes\left|\varphi\right\rangle _{t}\otimes\left|\phi\right\rangle _{aux}$, with $\left|\varphi\right\rangle _{c}=\alpha_{0}\left|0\right\rangle +\alpha_{1}\left|1\right\rangle $,
$\left|\varphi\right\rangle _{t}=\sum_{i=0,1}c_{i}\left|i\right\rangle $,
and some auxiliary state $\left|\phi\right\rangle _{aux}$. After
the controlled swap operation is applied (Fig. \ref{fig:cmeas}),
we find
\begin{equation}
\left|\Psi\right\rangle _{in}=\alpha_{0}\left|0\right\rangle _{c}\otimes\left|\varphi\right\rangle _{t}\otimes\left|\phi\right\rangle _{aux}+\alpha_{1}\left|1\right\rangle _{c}\otimes\left|\phi\right\rangle _{t}\otimes\left|\varphi\right\rangle _{aux}.\label{eq:cmeasini}
\end{equation}
Finally, a general projective measurement $\left\{ P_{0},P_{1}\right\} =\left\{ \left|\psi_{0}\right\rangle \left\langle \psi_{0}\right|,\left|\psi_{1}\right\rangle \left\langle \psi_{1}\right|\right\} $
is performed on the auxiliary qubit. In order to maintain the weights
of the superposition unchanged, the auxiliary qubit has to be suitably prepared
depending on the measurement basis and the target state. Therefore,
the measurement basis, as well as the amplitude probability distribution
of the target state, has to be known. The target
state can be written in the measurement basis $\left\{ \left|\psi_{0}\right\rangle ,\left|\psi_{1}\right\rangle \right\} $,
i.e. $\left|\varphi\right\rangle _{t}=\sum_{j,i}\left\langle \psi_{j}\left|i\right\rangle \right.\left|\psi_{j}\right\rangle $.
The auxiliary state has to be prepared with amplitude probabilities
$\left|\sum_{i}\left\langle \psi_{j}\left|i\right\rangle \right.\right|^{2}$,
also written in the measurement basis. In this case, after the measurement
is performed and the outcome, say $0$ (from $P_{0}$), is obtained,
the resulting global state is
\begin{equation}
\left|\Psi\right\rangle _{f}=\alpha_{0}\left|0\right\rangle _{c}\otimes\left|\varphi\right\rangle _{t}\otimes\left|\psi_{0}\right\rangle _{aux}+\alpha_{1}\left|1\right\rangle _{c}\otimes\left|\phi\right\rangle _{t}\otimes\left|\psi_{0}\right\rangle _{aux},\label{eq:cmeasfinal}
\end{equation}
with probability $\left|\sum_{i}\left\langle \psi_{0}\left|i\right\rangle \right.\right|^{2}$.
In this way, a superposition of the target state being measured or not is generated.

Although this construction might not seem very useful at this stage, interesting properties arise from it when e.g. the target state is part of a larger entangled state. Consider a system $a$ from an arbitrary entangled state where $a$ belongs to party $A$, who performs the controlled measurement. The state of $a$ is determined by its reduced density operator $\rho_{a}$. Given the same projective measurement as before, i.e.  $\left\{ \left|\psi_{0}\right\rangle ,\left|\psi_{1}\right\rangle \right\}$, the auxiliary qubit of $A$ has to be adequately prepared in some pure state $\rho_{aux}=\left|\phi\right\rangle _{aux}\left\langle \phi\right|$. The weights of this state are again chosen in order to ensure the same probabilities
for the measurement outcomes as for $\rho_{a}$,
i.e.
\begin{equation}
\left|\left\langle \phi\right|\left.\psi_{i}\right\rangle \right|^{2}=\text{tr}\left(P_{i}\rho_{a}\right),
\end{equation}
where $P_{i}$ defines each projector $\left|\psi_{i}\right\rangle \left\langle \psi_{i}\right|$ of the measurement. By writing the auxiliary state in the measurement basis, $\left|\phi\right\rangle _{aux}=\alpha_{0}\left|\psi_{0}\right\rangle +\alpha_{1}\left|\psi_{1}\right\rangle $,
one can immediately see that the weights $\alpha_{0},\alpha_{1}$
need to be chosen accordingly to the diagonal elements of the reduced
density operator $\rho_{a}$, also written in the measurement basis, such
that
\begin{equation}
\left|\alpha_{i}\right|^{2}=(\rho_{a}^{ii})_{\left\{\left|\psi_{0}\right\rangle,\left|\psi_{1}\right\rangle \right\}}.
\end{equation}
Once the state is prepared, the controlled swap is performed between $\rho_{a}$ and $\rho_{aux}$, followed by the measurement of the auxiliary system. Since both branches have the same probability distribution, the coherence of the final state is guaranteed. For instance, if the controlled measurement in the $\sigma_{x}$ basis is done on parts of a larger maximally entangled state, weights can always be kept equal by preparing the auxiliary system in the $\frac{1}{2}\mathbb{1}$ state. In the same direction, if the larger entangled state is some arbitrary graph state, this construction allows us to coherently cut qubit $a$, ending up with a superposition of the system $a$ being part ---or not--- of the graph state. We explain in detail this, and the completion of other controlled tasks, in the following section.

Note that we require partial knowledge of the measurement basis and the target state. However, this does not represent a problem for our purposes, as we show later. Note also that generalization to qudits and to general multi-outcome measurements is straightforward.

\subsection{Controlled measurements on parts of entangled states}
We introduce different tools where the mechanisms presented above allow
us to coherently control classical tasks for different purposes. We restrict the analysis, without loss of generality, to qubit systems and superpositions with two constituents for simplicity, but a generalization for an arbitrary number of elements and for qudit systems is straightforward. Crucially, the processes do not change the initial amplitudes in any case.

\subsubsection{Controlled sending}\label{sec:csend}
Consider the simplest scenario of sending quantum information via teleportation, where
a state is teleported by performing a Bell measurement between the
state that is teleported and one constituent of a Bell state. One can add
control to this process and create a coherent superposition of sending
and not sending the state information by applying the following procedure.

\begin{figure}[h!]
\includegraphics[width=\columnwidth]{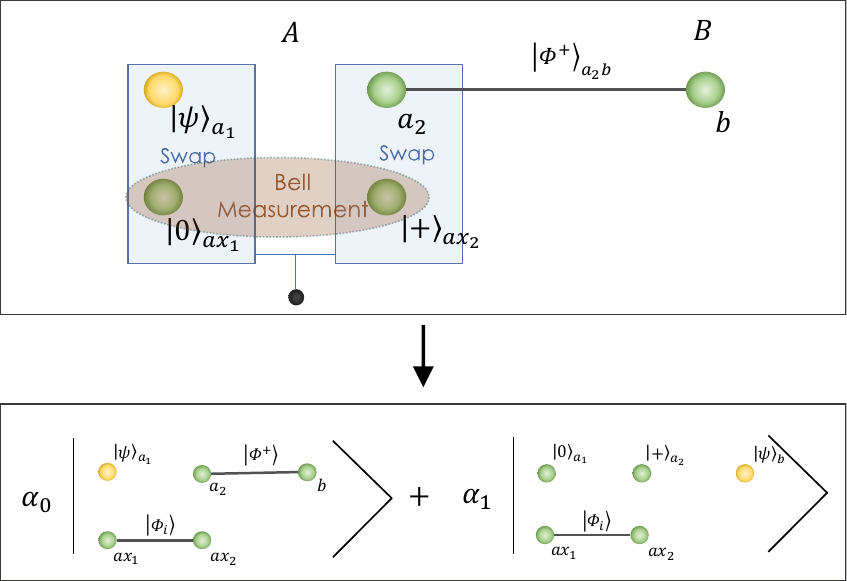}
\caption[h!]{\label{fig:csend} Schematic representation of controlled sending, both the initial state (top) and the final superposition (bottom). Party $A$ possesses four qubits which are coherently either swapped or not in a controlled and simultaneous way, followed by a Bell measurement of the auxiliary systems $ax_{1}, ax_2$. A superposition of the state $|\varphi\rangle$ teleported and kept by $A$ (i.e. not sent) is generated.}
\end{figure}

Consider the setting shown in Fig. \ref{fig:csend}. Parties $A$ and
$B$ initially share a Bell state and party $A$ prepares the arbitrary \textit{unknown}
state $\left|\psi\right\rangle _{a_{1}}$, which shall be teleported
in a controlled way. Additionally, party $A$ possesses two auxiliary
qubits, $ax_{1}$ and $ax_{2}$, initialized in the states $\left|0\right\rangle _{ax_{1}}$
and $\left|+\right\rangle _{ax_{2}}$ respectively. Party $A$ also
owns the control register $c$. We refer to Appendix B for details. The protocol involves the following steps. First, party $A$ applies
a controlled swap operation between qubits $a_{1},ax_{1}$ and
$a_{2},ax_{2}$ simultaneously, followed by a Bell measurement $\{|\Phi_i\rangle\}$ between qubits $ax_{1}$ and $ax_{2}$. The controlled swap transformation is described as

\begin{align}
C_{swap}&=\left|0\right\rangle _{c}\left\langle 0\right|\otimes\mathbb{1}_{a_{1}}\otimes\mathbb{1}_{ax_{1}}\otimes\mathbb{1}_{a_{2}}\otimes\mathbb{1}_{ax_{2}}+  \notag \\
& +\left|1\right\rangle _{c}\left\langle 1\right|\otimes U_{a_{1},ax_{1}}^{swap}\otimes U_{a_{2},ax_{2}}^{swap},\label{eq:cswap}
\end{align}
where $U_{i,j}^{swap}=\left|00\right\rangle \left\langle 00\right|+\left|01\right\rangle \left\langle 10\right|+\left|10\right\rangle \left\langle 01\right|+\left|11\right\rangle \left\langle 11\right|$.
More formally, the protocol starts with the following overall state:
\begin{equation}
\left|\Psi\right\rangle _{in}=\left(\alpha_{0}\left|0\right\rangle _{c}+\alpha_{1}\left|1\right\rangle _{c}\right)\left|\psi\right\rangle _{a_{1}}\left|\Phi^{+}\right\rangle _{a_{2}b}\left|0\right\rangle _{ax_{1}}\left|+\right\rangle _{ax_{2}},\label{eq:sendinginitial}
\end{equation}
where tensor products have been omitted for simplicity. After the controlled swap of Eq.~(\ref{eq:cswap}) is performed, a Bell measurement is carried out
on qubits $ax_{1}$ and $ax_{2}$. 
Assuming that the
outcome of the Bell measurement is $\left|\Phi_i\right\rangle _{ax_{1}ax_{2}},$
the final state reads
{\small{}
\begin{eqnarray}
\left|\Psi\right\rangle _{f}&=&\left(\alpha_{0}\left|0\right\rangle _{c} \left|\psi\right\rangle _{a_{1}} \left| \Phi^+\right\rangle _{a_{2} b}+\alpha_{1}\left|1\right\rangle _{c} \left|0\right\rangle _{a_{1}}\left|+\right\rangle _{a_{2}} \sigma_i\left|\psi\right\rangle _{b} \right) \nonumber\\
&&\left|\Phi_i\right\rangle _{ax_{1}ax_{2}}.\label{eq:sendingfinal}
\end{eqnarray}}Note that the global state has been re-normalized and the weights of the superposition remain unchanged. 
Depending on the measurements outcome $i$, correction operations are necessary. This just involves controlled Pauli correction operations $\sigma_i$ on qubit $b$, which only  need to be applied if the state was sent (i.e. control state is $|1\rangle$). Since the control register belongs to party $A$, this correction can also be implemented by $B$ always applying the unitary $\sigma_{i}$ on qubit $b$, followed by party $A$ applying the controlled unitary $\left|0\right\rangle_{c} \left\langle 0\right|\otimes\sigma_{i}^\intercal +\left|1\right\rangle_{c} \left\langle 1\right|\otimes\mathbb{1}$.
We refer the reader to Appendix B for details.

Observe that if the control qubit is in the state $\left|0\right\rangle$, the protocol preserves the state  $\left|\psi\right\rangle $ in the qubit $a_{1}$ of party $A$, and the Bell state $|\Phi^+\rangle$ shared between $A$ and $B$ (qubits $a_{2}$ and $b$). A coherent superposition of sending and not sending the state $\left|\psi\right\rangle$ 
is hence achieved. 
Note that we are able to teleport an unknown state, i.e. the restrictions of partially knowing the state on which applying the controlled measurement do not apply. The reason is that we perform the measurement on part of a maximally entangled state in one branch, which corresponds to a uniform probability distribution for the different measurement outcomes, independent of the state to be teleported. The same is true for the second branch, where the Bell measurement is performed on a product state $|0\rangle|+\rangle$ instead. In Sec. \ref{sec:extralevel} we show a procedure to deterministically detach the  control register of Eq. (\ref{eq:sendingfinal}), ensuring the orthogonality of the constituents of the remaining superposition. 

We have shown how controlled sending works by means of quantum teleportation. Nevertheless, this formalism is in principle extendible for sending of information through quantum channels, where a dummy state or a desired state is sent through the channel in a coherent controlled way.

\subsubsection{Controlled cutting of graph states} \label{sec:ccut}
Based on the same mechanisms, we discuss
here the possibility of controlled-cutting on parts of graphs states
in a coherent way. Consider a general graph state of the
form Eq.~(\ref{obs:vertexdecomp}). Assume one aims to
construct a state in superposition of the unaltered graph state  and the
graph state with qubit $a$ measured out in the computational basis. The party $a$ additionally
owns a control register, and  one auxiliary system  initially
prepared in the $\left|+\right\rangle_{aux} $ state. The initial state is hence given by
\begin{equation}
\left|\Psi\right\rangle _{in}=\left(\alpha_{0}\left|0\right\rangle _{c}+\alpha_{1}\left|1\right\rangle _{c}\right)\left|G\right\rangle \left|+\right\rangle _{aux}.
\end{equation}
We now imitate a controlled cutting of qubit $a$ of the graph state by first applying a controlled swap between the qubits $a$
and $aux$ (see Eq.~\ref{obs:vertexdecomp}), leading to \small{
\begin{multline}
\left|\Psi\right\rangle=\alpha_{0}\left|0\right\rangle _{c}\left|G\right\rangle\left|+\right\rangle _{aux} + \\
\frac{1}{\sqrt{2}}\alpha_{1}\left|1\right\rangle _{c}\left(\left|+\right\rangle _{a}\left|G/a\right\rangle \left|0\right\rangle _{aux}+\left|+\right\rangle _{a}\prod\sigma_{z}^{N_{a}}\left|G/a\right\rangle \left|1\right\rangle _{aux}\right).
\end{multline}}A single-qubit projective measurement is now performed on the auxiliary system in the computational
basis. The resulting state reads
\begin{align}
\left|\Psi\right\rangle _{f}=&\frac{1}{\sqrt{2}}\alpha_{0}\left|0\right\rangle _{c}\left|G\right\rangle\left|i\right\rangle _{aux} + \\  \notag
&\frac{1}{\sqrt{2}}\alpha_{1}\left|1\right\rangle _{c}\left|+\right\rangle _{a}\left(\prod\sigma_{z}^{N_{a}}\right)^{i}\left|G/a\right\rangle \left|i\right\rangle _{aux},
\end{align}
where $i=\{0,1\}$ is the measurement outcome and the state is not normalized.  Assuming the outcome of the measurement
is the $\left|0\right\rangle $ state, and after re-normalization, the final state is
\begin{equation}
\left|\Psi\right\rangle _{f}=\left(\alpha_{0}\left|0\right\rangle _{c}\left|G\right\rangle +\alpha_{1}\left|1\right\rangle _{c}\left|+\right\rangle _{a}\left|G/a\right\rangle \right)\left|0\right\rangle _{aux}.\label{eq:ccutt}
\end{equation}
Observe that the graph state $\left|G\right\rangle$ does not change if the control qubit is $\left|0\right\rangle$, but the operation removes all the edges between vertex $a$ and its neighbourhood of the graph state $\left|G\right\rangle$ if the control qubit is $\left|1\right\rangle$. Furthermore, the weights of the superposition remain unchanged.

In case the measurement outcome is $\left|1\right\rangle$, a controlled correction unitary of the form $\left|0\right\rangle \left\langle 0\right|\otimes\mathbb{1}+\left|1\right\rangle \left\langle 1\right|\otimes\prod_{N_{a}}\sigma_{z}$ is required to recover state Eq.(\ref{eq:ccutt}). Note that this operation is always realizable locally since the rest of devices, concretely $N_{a}$, possesses their own control systems, see Eq. (\ref{eq:requeststate}), before applying the final transformation of Eq. (\ref{eq:aftercontrolledunitaries}).

\subsubsection{Controlled merging and state manipulation} \label{sec:cmerg}
These techniques can be further combined and extended to obtain
full functionality for controlled state preparation of graph states. In order to obtain this functionality,
we discuss here controlled-merging of different graph states. 
\begin{figure}[h!]
\includegraphics[width=\columnwidth]{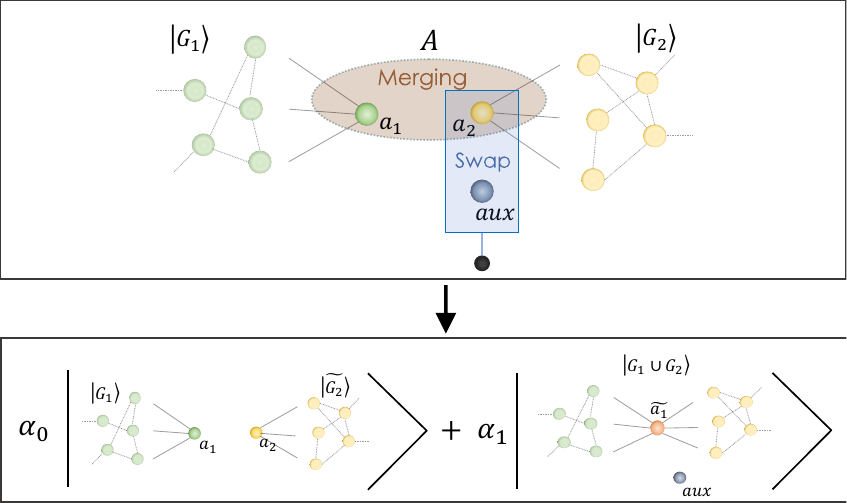}
\caption[h!]{\label{fig:cmerg} Schematic representation of controlled merging where the protocol is accomplished in the following way. If the control register is in the $\left|0\right\rangle $ state, the protocol preserves the two graph states unaltered, while the procedure merges the two graph states if the control register is in the $\left|1\right\rangle $ state.}
\end{figure}
Consider two graph states $\left|G_{1}\right\rangle ,\left|G_{2}\right\rangle $
of the form Eq.~(\ref{obs:vertexdecomp}), which we
want to merge the vertices $a_{1},a_{2}$. We make use of the merging measurement
operation defined by $\left\{ P_{0}=\left|0\right\rangle \left\langle 00\right|+\left|1\right\rangle \left\langle 11\right|,P_{1}=\left|0\right\rangle \left\langle 01\right|+\left|1\right\rangle \left\langle 10\right|\right\} $
(see Sec.~\ref{sec:backmerg}). Assume that qubits
$a_{1},a_{2}$ belong to party $A$, which also possesses an auxiliary
qubit prepared in $\left|+\right\rangle _{aux}.$ The
initial state is therefore $\left|\Psi\right\rangle _{in}=\left(\alpha_{0}\left|0\right\rangle _{c}+\alpha_{1}\left|1\right\rangle _{c}\right)\left|G_{1}\right\rangle \left|G_{2}\right\rangle \left|+\right\rangle _{aux}.$
The controlled merging comprises the following steps. First, party $A$ performs a controlled swap  between qubits $a_{2}$
and $aux$. Note that the swap operation is now applied in case the control register is in the $\left|0\right\rangle$ state. Next, party $A$ applies the merging measurement of Eq. (\ref{eq:mergedgraphs}) on qubits $a_{1}$
and $a_{2}$, merging them into one vertex $\tilde{a_{1}}$. Observe that the superposition amplitudes do not change. More precisely, the state before the merging measurement is:

\begin{align}
&\left|\Psi\right\rangle = \frac{1}{2}\alpha_{0}\left|0\right\rangle _{c}\left(\left|0\right\rangle _{a_{1}}\left|G_{1}/a_{1}\right\rangle +\left|1\right\rangle _{a_{1}}\prod\sigma_{z}^{N_{a_{1}}}\left|G_{1}/a_{1}\right\rangle \right)\cdot \notag \\
&\left(\left|+\right\rangle _{a_{2}}\left|G_{2}/a_{2}\right\rangle \left|0\right\rangle _{aux}+\left|+\right\rangle _{a_{2}}\prod\sigma_{z}^{N_{a_{2}}}\left|G_{2}/a_{2}\right\rangle \left|1\right\rangle _{aux}\right)+ \notag \\
&\hspace{0.75cm}+\alpha_{1}\left|1\right\rangle _{c}\left|G_{1}\right\rangle \left|G_{2}\right\rangle \left|+\right\rangle _{aux},
\end{align}
up to normalization. In Appendix C the details of the  process are provided.  If the merging measurement is now performed, and assuming that
the outcome is $0$, the resulting state reads 
\begin{equation}
\left|\Psi\right\rangle _{f}=\alpha_{0}\left|0\right\rangle _{c}\left|G_{1}\right\rangle \left|G_{2}\right\rangle+\alpha_{1}\left|1\right\rangle _{c}\left|G_{1}\cup G_{2}\right\rangle \left|+\right\rangle _{a_{2}},\label{eq:cmergfinal}
\end{equation}
where qubits $aux$ and $a_{2}$ have been relabelled (see Appendix C). In case the outcome $1$ is found in the merging measurement, a controlled correction operation of the form $\left|0\right\rangle_{c} \left\langle 0\right|\otimes\mathbb{1}+\left|1\right\rangle_{c} \left\langle 1\right|\otimes\prod\sigma_{z}^{N_{a_{2}}}$ is required  to recover the state Eq. (\ref{eq:cmergfinal}). Therefore, the final state consists of a superposition between two  graph states preserved unaltered, if the control qubit is in the $\left|0\right\rangle$ state, and two merged graph states, if the control register is in the $\left|1\right\rangle$ state. Again, the reason why the process can be performed on unknown (connected) graph states is that at least one of the qubits to be measured is part of a maximally entangled state.

\section{Example of quantum controlled request. Orthogonality of states} \label{sec:qcrequest}

In this section we provide a detailed example of a particular quantum controlled request in a quantum network. As mentioned before, orthogonality of constituents of the final superposition turns out to be a crucial property to guarantee coherence when dispatching the control register. We discuss possible solutions to tackle this problem in general.

\subsection{Example}\label{sec:example}

\begin{figure}[h!]
\includegraphics[width=\columnwidth]{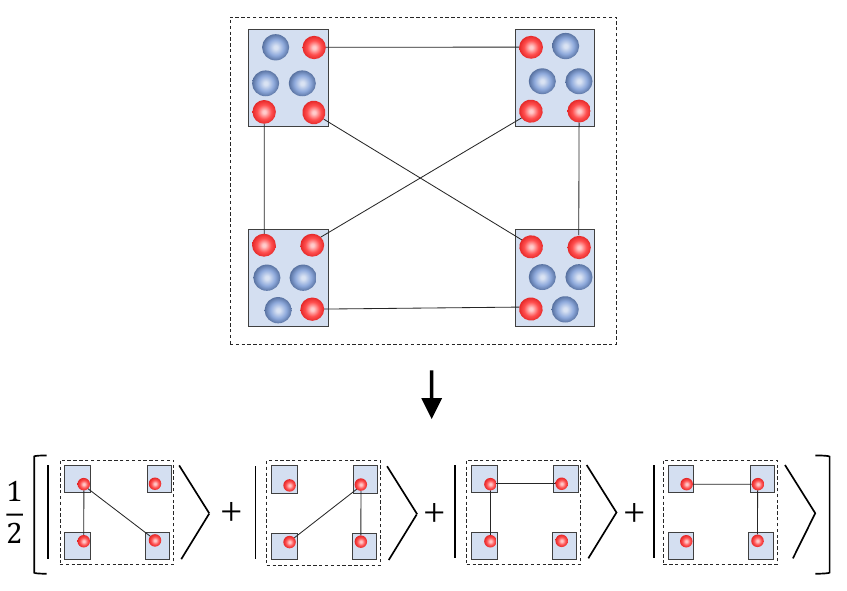}
\caption[h!]{\label{fig:example} Schematic example of the initial state (upper figure) of Eq.~(\ref{eq:bellsinitial}) and final desired superposition (bottom figure) of Eq.~(\ref{eq:extralevel2}). Red vertices represent the resource qubits and blue vertices represent the auxiliary qubits, which are omitted in the bottom picture for illustrative simplicity.}
\end{figure}

With the tools introduced in the previous section,
multiple tasks can be completed in a coherent controlled manner. Consider
the following example of a network request. Each device of the network represents
a single user or node, and each of them shares a Bell pair with each of the remaining ones. This is the initial entangled resource state in an entanglement-based top-down approach to quantum networks. Every
node possesses an additional auxiliary qubit per each resource Bell state it owns, apart from the control register. In principle also a full quantum description of the desired actions can be provided by adding a program register. In total,
each device stores $2+2(n-1)$ qubits, where $n$ is the number of users. For simplicity, we assume that the network
consists only of four devices, such that the resource state is given
by (see Fig. \ref{fig:example})
\begin{equation}
\left|\psi\right\rangle_{res} =\bigotimes_{(i,j) \in E}\left|\Phi^{+}\right\rangle _{ij},\label{eq:bellsinitial}
\end{equation}
with $\left|\Phi^{+}\right\rangle$ the Bell state of Sec. \ref{sec:background:bell} and $E=\{(i,j)|(1 \leq i < j \leq 4)\}$. 
Additionally, the auxiliary qubit of each device is initialized in the $\left|+\right\rangle $ state. Note that a Bell state is LU-equivalent to a graph state, and the choice of the state of the auxiliary qubits is hence motivated by the merging measurement we consider (see Sec. \ref{sec:cmerg}) for graph states. In particular, the merging of a qubit of a graph state with a qubit in the $\left|+\right\rangle $ state, retains the graph state unchanged. Besides, weight invariance is also guaranteed for the superposition.

The desired target state, see Eq. (\ref{eq:targetstate}), is a equal-weight superposition of all the possible 3-party
combinations of GHZ states, i.e.
\begin{equation}
\left|\Psi\right\rangle _{f}=\frac{1}{2}\sum_{i=1}^{4}\left|i\right\rangle _{c}\left|0\right\rangle _{i}\left|\mathrm{GHZ}\right\rangle_{N/i}.\label{eq:ghzfinal}
\end{equation}
where $\left|\mathrm{GHZ}\right\rangle _{N/i}$ indicates a $\mathrm{GHZ}$ state shared among the other three parties, excluding system $i$. The remaining auxiliary states are omitted for simplicity.

The whole process is carried out by effective controlled operations
in every node. In each site, three "rounds" of controlled-merging operations (see Sec.~\ref{sec:cmerg}) are performed. Each round specifically consists in taking two-by-two resource qubits, that are subsequently merged ---or not--- in a controlled way. Following Sec. \ref{sec:cmerg}, the controlled swap
operations of each round are defined by the request control register
of each of the devices. In each round the result can be the merging of the two resource Bell states at that site, or the resource Bell pair not becoming part of a larger GHZ state, in case the swap has been performed. In this last situation, the Bell state is retained at the auxiliary level. Observe that for retaining the Bell pairs at the auxiliary level, synchronized actions between devices are required. In particular, in both devices the swap operation need to be performed within a particular branch of the superposition, such that the initial resource Bell state they share is kept between their auxiliary systems. In order to obtain the target state of Eq.~(\ref{eq:ghzfinal}), one can combine
controlled-merging with controlled-cutting (Sec. \ref{sec:ccut}), such that the auxiliary qubits are measured after each round and no entanglement is kept at the auxiliary level.

Consider now only {\it one} branch of the superposition, the corresponding to the element $i=1$ of Eq.~(\ref{eq:ghzfinal}). Consider for instance the device $j=4$, which performs
three controlled-merging rounds, from which one involves the swapping, synchronized with device
$j=1$. The corresponding resource Bell state shared by $j=1$ and $j=4$ is therefore kept at the auxiliary level. The other two parties do not swap, and hence the resource Bell pairs are merged into a larger GHZ state (see Sec. \ref{sec:cmerg}), leading to
\begin{equation}
\left|\Phi^{+}\right\rangle _{14}^{(aux)}\left|\mathrm{GHZ}\right\rangle _{234},\label{eq:ghz1}
\end{equation}
where we have omitted the remaining resource and auxiliary qubit states for simplicity.
Observe that the Bell state in Eq. (\ref{eq:ghz1}) involves the auxiliary qubits of devices $j=1$ and $j=4$. We can combine this with a controlled cutting process (see Sec. \ref{sec:ccut}). After each merging round, the auxiliary qubit is hence always measured in the computational basis, such that the resulting state reads
\begin{equation}
\left|0\right\rangle _{1}\left|\mathrm{GHZ}\right\rangle _{234},\label{eq:ghz2}
\end{equation}
where we have assumed the outcome $\left|0\right\rangle $ of the
measurement is found and the remaining resource and all auxiliary qubit states are again omitted. Note again that merging and cutting affects only this branch of the superposition. In extension, this process applies for the different branches and the different network devices, such that the process of Fig.~\ref{fig:example} is accomplished.

Observe that, when combining
controlled merging and controlled cutting, only one auxiliary
qubit is required to be stored per site, independently of the resource state. This auxiliary
qubit is measured in each round and prepared again in the $\left|+\right\rangle $ state, such that it can be used for the next merging round. Note also that, in the case the controlled cutting is included, entanglement kept at the auxiliary level is destroyed. In contrast, if controlled cutting is not included in the procedure, the resource entanglement is kept anyway, either at the target or auxiliary level, but extra storage for auxiliary
qubits is needed for each device.

Following with the example, the state after the remaining controlled operations
and considering  the different branches is exactly of the form of Eq.~(\ref{eq:ghzfinal})  (see Fig.~\ref{fig:example}). Observe that the effect of unwanted measurement outcomes can be adequately
corrected locally, as explained in Sec.~\ref{sec:superptasks}.  These procedures are extensible for different initial entanglement resources and for different desired states within the superposition, including e.g. generation of superposition of states with
different entanglement properties. The applied operations  depend entirely on the desired target states.

In addition, one might aim to detach the control register from Eq. (\ref{eq:ghzfinal}), in order to obtain

\begin{equation}
\left|\Psi\right\rangle _{f}=\frac{1}{2}\sum_{i=1}^{4}\left|0\right\rangle _{i}\left|\mathrm{GHZ}\right\rangle_{N/i}, \label{eq:ghznocontrol}
\end{equation}
in analogy to Eq.~(\ref{eq:targetstate2}). From Eq.~(\ref{eq:ghzfinal}), the initiator  device has to measure its
control register  in the generalized Pauli $X$ basis.
However, two problems arises at this stage. First, one can see that
the states of the superposition of Eq.~(\ref{eq:ghznocontrol}) are not orthogonal
to each other. Therefore, Eq.~(\ref{eq:ghznocontrol}) is not a coherent superposition of
the different constituents. On the other hand, the $X$ measurement of
the control register can lead to unwanted phases in the different
elements of the superposition, which cannot be a-priori corrected.
This two inconveniences can be adequately overcome by considering the
following modification.

\subsection{Orthogonality of target states. Extra-level modification}\label{sec:extralevel}
We show a modification that allow us to go from Eq.~(\ref{eq:ghzfinal}) to Eq.~(\ref{eq:ghznocontrol}) deterministically, such that the coherence of the final superposition is guaranteed. This modification can be extended to more general cases.

A simple trick can be used to ensure orthogonality of the target
states. It consists in providing
extra levels (qudits) for certain systems and applying the controlled operation
\begin{equation}
C_{X}=\left|0\right\rangle _{c}\left\langle 0\right|\otimes\mathbb{1}_{i}+\left|1\right\rangle _{c}\left\langle 1\right|\otimes X_{i}^{2},\label{eq:controlextralevel}
\end{equation}
invoking the map  $\left\{ \left|0\right\rangle ,\left|1\right\rangle \right\} \rightarrow\left\{ \left|2\right\rangle ,\left|3\right\rangle \right\} $ between subspaces.
In Eq.~(\ref{eq:controlextralevel}), $X$ is the generalized Pauli operator for $d$-level systems,
such that $X\left|j\right\rangle =\left|j\oplus1\right\rangle $,
with addition mod $d$. This controlled operation becomes now part
of the procedure, implemented in suitable
places for the different branches. Observe that this operation should be applied before the control registers, except the initiator one, are measured (see Sec. \ref{sec:superptasks}). Following the example of the previous section,
the final state with the extra-level modification reads now
\begin{equation}
\left|\Psi\right\rangle _{f}=\frac{1}{2}\sum_{i=1}^{4}\left|i\right\rangle _{c}\left|2\right\rangle _{i}\left|\mathrm{GHZ}\right\rangle_{N/i}.\label{eq:extralevel1}
\end{equation}
The control register can be now measured, ensuring the
orthogonality of the elements of the resulting superposition. Not only the orthogonality issue is resolved, but one can also correct  the state in case of
unwanted phases come from the measurement of the control register.
Depending on the outcome of the measurement, each party applies an
adequate controlled correction unitary of the form $\left(\mathbb{1}-\left|2\right\rangle \left\langle 2\right|\right)\otimes\mathbb{1}+\left|2\right\rangle \left\langle 2\right|\otimes U(i)$,
where the operator $U(i)$ depends on the outcome of the measurement
and the position it is applied. A coherent superposition
\begin{equation}
\left|\Psi\right\rangle _{f}=\frac{1}{2}\sum_{i=1}^{4}\left|2\right\rangle _{i}\left|\mathrm{GHZ}\right\rangle_{N/i},\label{eq:extralevel2}
\end{equation}
is finally found, where all the constituents are now orthogonal to
each other. Note that before each device was required to store one
qudit system, corresponding to the request register. With this modification,
more than one d-level systems have to be stored in each station. We remark that two
qubit systems can be always be embedded to play the role of a four dimensional
system, by defining $\left\{ \left|00\right\rangle ,\left|01\right\rangle ,\left|10\right\rangle ,\left|11\right\rangle \right\} \rightarrow\left\{ \left|0\right\rangle ,\left|1\right\rangle ,\left|2\right\rangle ,\left|3\right\rangle \right\} $.

This modification can also be implemented,
for instance, in the controlled sending setting introduced in Sec.~\ref{sec:csend},
where target states are not orthogonal if the control register is
measured. There, one just need to implement, at the final stage of the process, the transformation $\left\{ \left|0\right\rangle ,\left|1\right\rangle \right\} \rightarrow\left\{ \left|2\right\rangle ,\left|3\right\rangle \right\} $ in a controlled way. From Eq. (\ref{eq:sendingfinal}), this controlled operation is applied
on the auxiliary system $a_{2}$ if the control register is in the $\left|0\right\rangle_{c}$ state, and on the qubit $a_{1}$ if the control register is in $\left|1\right\rangle_{c}$

This mechanism can also  be extended and used in more general situations, for those target states whose constituents are not orthogonal. In order to be able to apply this modification, for each target state in the superposition there should always be a different node which is not involved in the process.

\section{Quantum controlled addressing} \label{sec:addressing}

So far, we have shown how to provide quantum networks with a truly quantum functionality, based on the generation of superpositions of different tasks in a controlled way. This is accomplished by mimicking the behavior of certain classical tasks in a controlled way, always performing the task.

\begin{figure}[h!]
\includegraphics[width=\columnwidth]{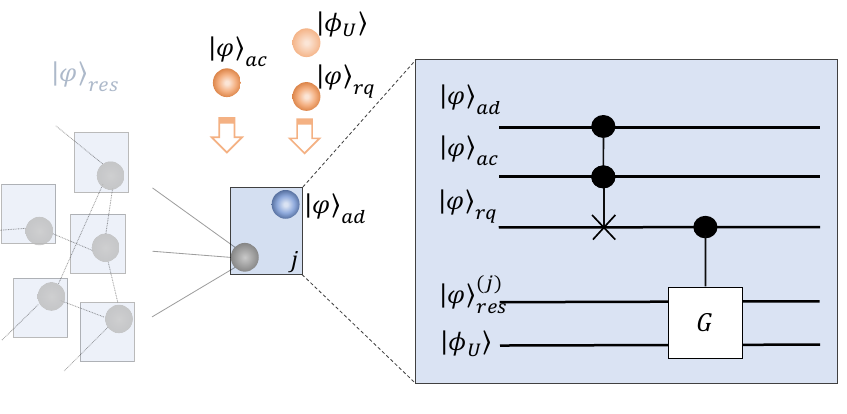}
\caption[h!]{\label{fig:ctoff} Schematic illustration of controlled addressing. Each device of the network possesses an addressing register prepared in $\left|\varphi\right\rangle_{ad}$, as well as the resource and auxiliary qubits (grey vertices). Note that each device owns some number of resource and auxiliary qubits, but we represent all of them unified in one for  simplicity. Additionally, each device is provided with a request register $\left|\varphi\right\rangle_{rq}$, a program register $\left|\phi_{U}\right\rangle$ and an activation register $\left|\varphi\right\rangle_{ac}$. A  Toffoli gate is applied, such that only if the addressing and activation registers match, the device is effectively turned on, and the control of the operations is now defined by the request register state. A programmable operation $G$ (Eq.~\ref{eq:gprog}) is applied, controlled by the request register, resulting in the coherent application of some unitary $U$ on the device's resource qubits $\left|\varphi\right\rangle^{(j)}_{res}$ (see Sec.~\ref{sec:fullyq} for details).}
\end{figure}

An additional addressing feature can be included on top of our approach. Consider again a quantum network consisting of different devices. Each device would own an addressing system, adequately prepared in some quantum state. This register works as an addressing register ($ad$), identifying each of the devices. Additionally, depending on the target network objective, they are provided with an activation register ($ac$). The addressing and activation registers determine, when compared, if the device takes an active role in the process or not.  The comparison is simply performed by a generalized Toffoli operation \cite{Daboul2003}, acting locally on each device $j$ (see Fig. \ref{fig:ctoff}), i.e.

\begin{equation}
\mathrm{T}^{(j)}=\sum_{p,s=0}^{d-1}\delta_{p,s}P_{p}^{ad}\otimes P_{s}^{ac}\otimes \left(X^{rq}\right)^{s},\label{eq:multitoffoli}
\end{equation}
where $P_{i}^{m}=\left|i\right\rangle \left\langle i\right|$ are
the projectors of the $m$ register, and $X$ is the generalized Pauli operator $\sigma_{x}$ of dimension $d$. The target state where the operation acts on is the request register ($rq$) ---also denoted as control register ($c$) in previous sections---, that determines the control over the actions that the device applies in order to generate the requested superposition. The request system is hence prepared in such a way that no actions are invoked, in a controlled way, if the Toffoli operation is unsuccessful, and some particular operations are invoked if the Toffoli is performed, taking into account the effect of the Toffoli operation in each device.

In summary, the roles of the involved registers are the following.  The addressing qubit labels or addresses each network device. The activation qubit identifies each device  unequivocally and decides its activation. Only if these two register states match, the device is correctly identified and effectively turned on. Otherwise, in case their states differ, it means that the device is not involved in the process for that concrete request, and there is no need to activate it.  This process is only implemented once. In case the device is activated, the role of control register is thus transferred to the request register. Subsequently, this request register defines the control of the applied operations (see Fig. \ref{fig:ctoff}). The information of the actions requested to obtain the desired output state is encoded in the program register.

Following the fully quantum setting description of Eq.~(\ref{eq:global1}), in case the device is effectively activated,
the controlled-operation of Fig. \ref{fig:ctoff} invokes an operation
of the form $\prod_{k}\left|k\right\rangle_{rq} \left\langle k\right|\otimes U_{k}^{(j)}$ via the program register, as explained in Sec.~\ref{sec:fullyq}. The rest of the procedure takes place in an analogous way than for previous sections, but with the additional controlled addressing feature, which completes the fully quantum description of Sec.~\ref{sec:fullyq}. Observe that the inclusion of the activation register, together with the program register, allows in some way for a remote invocation of the applied controlled unitary operations.

\section{Applications and features}\label{sec:examples}

In this section we analyze the applicability of the mechanisms and
techniques presented so far. We have introduced procedures to obtain coherent superpositions of tasks. We discuss
now how the generation of these superpositions can lead to advantages  in different contexts.
In order to do a reliable analysis that justifies the benefit of generating such superposed states (see Eq. \ref{eq:targetstate2}), in general, one has to compare them, firstly, with the corresponding classical
mixtures, defined by the a-priori classical probability distribution
of the different target states, i.e. $\left|\psi_{T}\right\rangle =\sum_{i}\alpha_{i}\left|\psi_{i}\right\rangle \left\langle \psi_{i}\right|$. Secondly, one has to compare the superposition features with the properties of each individual constituent generated according to some probability distribution $\left\{ \alpha_{i},\,\left|\psi_{i}\right\rangle \right\}
$. Orthogonality between the different
constituents of the final superposition is a crucial requirement
to make these comparisons trustworthy.

For simplicity, we assume superpositions with equal-weight coefficients in most of the examples. However, they are extendible to arbitrary-weight superpositions.

\subsection{Entanglement structure}

An important advantage that motivates the generation of states in
superposition is the entanglement structure it provides. We show two  examples where this structure is found beneficial for the generated superposed states.

\subsubsection{Bound vs. maximal entanglement}

Consider a network of four devices (see e.g. Fig.~\ref{fig:example})
which implements the protocols explained in previous sections in order
to generate the state
\begin{equation}
\left|\psi\right\rangle =\frac{1}{2}\sum_{i=0}^{3}\left|i\right\rangle _{c}\left|\Phi_{i}\right\rangle _{12}\left|\Phi_{i}\right\rangle _{34},\label{eq:smolin1}
\end{equation}
where $\left\{\Phi_{i}\right\}$ defines the basis of Bell states
for qubits (Eq.~\ref{eq:bellbasis}). Note that in this case, the control register can be deterministically
detached without changing the weights or turning into extra-level systems,
by just measuring it in a generalized $\sigma_{x}$ basis. The corresponding phases of
the different measurement outcomes can be corrected in a local way
by appropriate unitary transformations of the form $\left\{ \mathbb{1},\sigma_{z}^{1}\sigma_{z}^{2},\sigma_{x}^{1}\sigma_{x}^{2},\sigma_{x}^{1}\sigma_{x}^{2}\sigma_{z}^{1}\sigma_{z}^{2}\right\} $,
such that the resulting state reads
\begin{equation}
\left|\psi\right\rangle =\frac{1}{2}\sum_{i=0}^{3}\left|\Phi_{i}\right\rangle _{12}\left|\Phi_{i}\right\rangle _{34}.\label{eq:smolin2}
\end{equation}
Note that the corresponding classical mixture is $\rho=\frac{1}{4}\sum_{i=0}^{3}\left|\Phi_{i}\right\rangle _{12}\left\langle \Phi_{i}\right|\otimes\left|\Phi_{i}\right\rangle _{34}\left\langle \Phi_{i}\right|$,
the well-studied state so-called Smolin state \cite{Smolin2001}. An important
property of this state is that it is bound entangled, meaning that,
although entangled, the entanglement of distillation is zero, and hence
pure entanglement cannot be locally created between any bipartition of a single qubit
and any of the others. In opposition, the state Eq.~(\ref{eq:smolin2})
is purely maximally entangled with respect to all the bipartitions, i.e. the reduced density operator of each bipartite cut corresponds to the identity operator.

Additionally, when comparing the superposed state of Eq.~(\ref{eq:smolin2}) with the individual constituents chosen from a given probability distribution, one can clearly see that, while all bipartitions  of the superposed state are maximally entangled, there exist bipartitions  with zero entanglement, e.g. between systems $\{1,2\}$ and $\{3,4\}$, for the individual case.
Therefore, there exists a clear motivation in terms of entanglement for generating states of the form Eq.~(\ref{eq:smolin2}), instead of working with the individual elements or the classical mixture of the constituents.

\subsubsection{Entanglement vs. no-entanglement} \label{Sec:entvsno}

Our approach can also be conceived as a hierarchical entanglement decision
tool, where the control register can determine, at a later stage,
if the final state is entangled or not. This can be simply seen from
the following example. With the tools introduced in this work, we can generate a state of the form
\begin{equation}
\left|\psi\right\rangle =\frac{1}{\sqrt{2}}\sum_{i=0}^{1}\left|i\right\rangle _{c}\left|ii\right\rangle _{12},
\end{equation}
between two systems $1$ and $2$. Subsequently, the initiator device, which owns the control system, can decide from the outside, by appropriately choosing the measurement basis, if the final
state will be entangled or not. A projective measurement of the control qubit in the Pauli $\sigma_{x}$ basis leads to a Bell state between parties $1$ and $2$, while a measurement in the Pauli $\sigma_{z}$ basis leads to a product state, up to local corrections. In both cases, these corrections can be applied locally. In particular, for the first situation, the state $\ket{\Phi^{+}}$ can be obtained deterministically.

\subsection{Stability under losses }

In several cases, states in superposition also exhibit
stronger stability, in terms of entanglement, under errors or losses.

\subsubsection{Superposition of GHZ states}

Consider again the state Eq.~(\ref{eq:extralevel2}), i.e.
\begin{equation}
\left|\Psi\right\rangle _{f}=\frac{1}{2}\sum_{i=0}^{3}\left|2\right\rangle _{i}\left|\mathrm{GHZ}\right\rangle _{N/i},\label{eq:ghzsuperpos}
\end{equation}
which consists of all the possible permutations of $3$-party $\mathrm{GHZ}$
states shared among $4$ different parties, where the extra-level
modification is included to ensure orthogonality between elements. In Sec.~\ref{sec:example} and Sec.~\ref{sec:extralevel} the detailed process to generate this state is provided. When e.g. two systems of the state Eq.~(\ref{eq:ghzsuperpos}) are lost or traced out, the remaining state is still entangled at a bipartite level. More precisely, the remaining state has negativity $N=0.1$, where the negativity is defined
as the absolute value of the sum of the negative eigenvalues of the partial transposition
of the state, $N=\frac{\left|\left|\rho^{\mathrm{T_{a}}}\right|\right|-1}{2}$. However, if one studies the corresponding classical mixture, i.e.
\begin{align}
\frac{1}{4}\sum_{i=0}^{3}\left|2\right\rangle _{i}\left\langle 2\right|\otimes\left|\mathrm{GHZ}\right\rangle _{N/i}\left\langle\mathrm{GHZ}\right|,\label{eq:ghzmixture}
\end{align}
under analogous circumstances, i.e. two systems are lost, one can observe that the state Eq.~(\ref{eq:ghzmixture}) becomes separable. In the same direction, in case only one system is lost, the state resulting from Eq.~(\ref{eq:ghzsuperpos}) presents larger entanglement than the one resulting from the classical mixture, Eq.~(\ref{eq:ghzmixture}) with respect to all the bipartitions.

One can also compare the superposition of Eq. (\ref{eq:ghzsuperpos}) with respect to the individual constituents generated from a probability distribution. In this case, given a single $\mathrm{GHZ}$ state, two immediate features arise. First, as before, when two systems are lost, it always leads to a separable state. Secondly, assume the control register has not been detached and one system is lost. Therefore, from the superposition of Eq. (\ref{eq:ghzsuperpos}) one can always probabilistically go to a perfect GHZ shared by three parties, by simply measuring the control register in the Pauli $\sigma_{z}$ basis. This is not possible in the individual-constituent case.

Similar results are found when considering superposition of all the possible Bell pair connections between three parties, as well as when considering superposition states of increasing entanglement order, i.e. between product states, Bell states and GHZ states as elements of the superposition.

\subsubsection{Superposition of states with different entanglement structures}
One can also provide the control register with the possibility of deciding the entanglement structure of the final state. Similarly to Sec. \ref{Sec:entvsno}, we can generate a state of the form

\begin{equation}
\left|\psi\right\rangle =\frac{1}{\sqrt{2}}\left(\left|0\right\rangle_{c}\left|\mathrm{GHZ}_{1}\right\rangle _{1234}+\left|1\right\rangle_{c}\left|\mathrm{GHZ}_{2}\right\rangle _{1234}\right),
\end{equation}
where $\left|\mathrm{GHZ}_{1}\right\rangle=\frac{1}{\sqrt{2}}(\left|0000\right\rangle-\left|1111\right\rangle)$ and $\left|\mathrm{GHZ}_{2}\right\rangle=\frac{1}{\sqrt{2}}(\left|0011\right\rangle+\left|1100\right\rangle)$. Therefore, by choosing the measurement basis of the control register, the entanglement structure of the final state is decided. If the control register is measured in the Pauli $\sigma_{z}$ basis, a $\mathrm{GHZ}$ state is found. However, if the control  qubit is measured in the Pauli $\sigma_{x}$ basis, a cluster state $C_{1D}$ is obtained (up to phase corrections), where $\left|C_{1D}\right\rangle =\frac{1}{2}\left(\left|0000\right\rangle +\left|0011\right\rangle +\left|1100\right\rangle -\left|1111\right\rangle \right)$
is the one dimensional cluster state.

In a similar direction, one can also generate superposition of states
with different entanglement structure. Consider for instance the state
\begin{equation}
\left|\psi\right\rangle =\frac{1}{\sqrt{2}}\left|0\right\rangle _{c}\left|\mathrm{GHZ}\right\rangle _{1234}+\frac{1}{\sqrt{2}}\left|1\right\rangle _{c}\left|C_{1D}\right\rangle _{1234},\label{eq:ghzcluster}
\end{equation}
Note that the states $C_{1D}$ and $\mathrm{GHZ}$ are orthogonal.  The state of Eq. (\ref{eq:ghzcluster}) exhibits similar properties as the ones in the previous examples. Once the control register is measured and detached, since orthogonality condition is fulfilled, and considering that
two systems are lost, the remaining state is still
entangled, with negativity $N=0.35$ for each of the branches of the measurement.
In opposition, the corresponding classical mixture,
\begin{equation}
\left|\psi\right\rangle =\frac{1}{2}\left|\mathrm{GHZ}\right\rangle _{1234}\left\langle \mathrm{GHZ}\right|+\frac{1}{2}\left|C_{1D}\right\rangle _{1234}\left\langle C_{1D}\right|,
\end{equation}
becomes separable under the same assumptions. For one lost system, the remaining entanglement is again larger in the superposition case, in terms of negativity,  for any bipartition cut. Similarly, if one considers each state individually, one finds that each state becomes separable if two systems are lost, while the superposition remains entangled.

Further examples exist where the superposition always exhibits stronger entanglement robustness under errors or losses of systems, when comparing with the corresponding classical mixture. In particular, the examples presented above stress the motivation of generating superposition
states due to their entanglement stability under losses, where indeed the corresponding classical mixtures become separable in some situations.

\subsection{Superposed destinations}

The techniques introduced in Sec.~\ref{sec:csend} to add control to sending of quantum information can be used to distribute information to several parties within a network in a coherent way. Consider an scenario where the initiator device shares a Bell state with each of the network devices. Given some arbitrary state $\left|\varphi\right\rangle=\alpha\left|0\right\rangle +\beta\left|1\right\rangle$ that the initiator owns, and given the tools of Sec.~\ref{sec:csend}, one can directly generate a state of the form
\begin{equation}
\left|\varPsi\right\rangle =\frac{1}{\sqrt{n}}\sum_{i}^{n}\left|i\right\rangle _{c}\left|\varphi\right\rangle _{i}\left|0\right\rangle^{\otimes n-1}_{n\neq i}.
\end{equation}
One finds a superposed state, whose elements correspond to the state $\left|\varphi\right\rangle$ being distributed among each of the network devices. In order to detach the control register, one can again apply the controlled-extra level modification explained in Sec.~\ref{sec:extralevel}, which allow us to correct unwanted phases and, besides, make the final state constituents orthogonal to each other. The final state then reads
\begin{equation}
\left|\varPsi\right\rangle =\frac{1}{\sqrt{n}}\sum_{i}^{n}\left|\varphi\right\rangle _{i}\left|2\right\rangle _{i\oplus1}\left|0\right\rangle^{\otimes n-1}_{n\neq i,i\oplus1},
\end{equation}
where $\oplus$ denotes addition mod~$n$. Note that unwanted phases from the measurement of the control register can be suitably corrected by applying controlled operations of the form $\left|2\right\rangle\left\langle 2\right|\otimes U^{(i)}$, where $U^{(i)}$ depends on the position of the $\left|2\right\rangle$ state.

An intimidate application of this example is the case where one aims to send information to a subset of parties, but does not want to decide yet who will ultimately receive the information.

\subsection{Superposition of paths}

\begin{figure}[h!]
\includegraphics[width=\columnwidth]{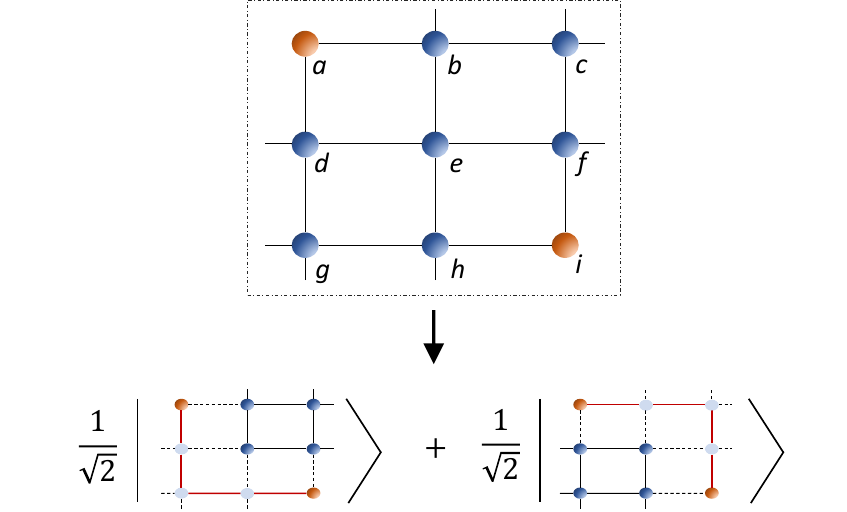}
\caption[h!]{\label{fig:paths} Example of superposition of paths. Given some resource state (e.g. 2D cluster state) connecting several nodes in a network, a coherent superposition can be obtained in a controlled way between different network routes in order to connect two parties ($a$ and $i$).}
\end{figure}

One can also understand the mechanisms for generating states in superposition
in the following way. Consider a network with several nodes connected
by some resource state (e.g. a 2-D cluster state), where direct communication links between network constituents aim to be established. The simplest scenario, illustrated in Fig.~\ref{fig:paths}, consist of two parties, $a$ and $i$, between which a direct communication link should be provided. Different approaches for solving this, and more advance routing problems have been studied \cite{
Meter2013b,Pirandola16,Hahn2019}. Typically, for qubit graph state resources, the easiest way
to connect parties $a$ and $i$ (Fig.~\ref{fig:paths}) consists in finding the shortest path between the two parties and measure
the intermediate nodes in the Pauli $\sigma_{x}$ basis, followed by Pauli $\sigma_{z}$ measurements of the path neighbourhood. Following our approach, consider some resource state $\left|\psi\right\rangle _{res}$ and a control register attached, as explained in previous sections,
\begin{align}
\left|\Psi\right\rangle _{in}=\frac{1}{\sqrt{2}}\left(\left|0\right\rangle _{c}+\left|1\right\rangle _{c}\right)\left|\psi\right\rangle _{res}.\label{eq:paths1}
\end{align}
Based on the example of Fig.~\ref{fig:paths}, the desired target state is $\left|\Phi^{+}\right\rangle _{ai}$, and an equal-weight  superposition
between two different possible paths to reach that target state can be prepared, i.e.
\begin{align}
\left|\Psi\right\rangle _{in}=&\frac{1}{\sqrt{2}}\left|0\right\rangle _{c}\sigma_{x}^{d}\sigma_{x}^{g}\sigma_{x}^{h}\prod \sigma_{z}^{N_{0}}\left|\psi\right\rangle _{res}+  \notag \\
&\frac{1}{\sqrt{2}}\left|1\right\rangle _{c}\sigma_{x}^{b}\sigma_{x}^{c}\sigma_{x}^{f}\prod \sigma_{z}^{N_{1}}\left|\psi\right\rangle _{res}.\label{eq:paths2}
\end{align}
The Pauli operators $\sigma_{z}$ act on the neighbourhood of the elements of each path, where paths are labelled as $0$ and $1$, in order to isolate the desired Bell pair connection. Each constituent of the superposition defines the actions to be done for each case, leading to a final state of the form
\begin{align}
\left|\Psi\right\rangle _{in}=\frac{1}{\sqrt{2}}\left|0\right\rangle _{c}\left|\Phi^{+}\right\rangle _{ai}\left|\psi{'}\right\rangle _{res}+\frac{1}{\sqrt{2}}\left|1\right\rangle _{c}\left|\Phi^{+}\right\rangle _{ai}\left|\psi{''}\right\rangle _{res},\label{eq:paths3}
\end{align}
where the desired Bell pair is obtained in both cases, but following different directions, therefore modifying the resource state in different ways.
One possible application of this scheme regards the protection against errors. Since a path does not
have to be chosen beforehand, the process is consequently protected
against possible failures or errors of individual nodes of the resource.

\subsection{Encoding and delocalized quantum information} \label{sec:encod}

One can also conceive the quantum network as a tool to distribute and store quantum information. In particular, the protocols allow one to generate the states required in order to encode quantum information into a network of several devices in a straightforward way. Instead of getting rid of the control register, it can be used to encode and delocalize quantum information. Following
\cite{delocalizedinfo}, consider a state of the form
\begin{equation}
\left|\psi\right\rangle =\frac{1}{\sqrt{2}}\left(\left|0\right\rangle _{c}\left|0_{L}\right\rangle +\left|1\right\rangle _{c}\left|1_{L}\right\rangle \right),\label{eq:encoding}
\end{equation}
with two codewords $\left|0_{L}\right\rangle ,\left|1_{L}\right\rangle$,
which are essentially two orthogonal states of $n$ qubits correpsonding to some error correction code. As
we have shown, the tools presented in this paper provide the possibility of generating states of the form Eq.~(\ref{eq:encoding}), for different codeword pairs. From
state Eq.~(\ref{eq:encoding}), the control system can encode the information of any arbitrary
state $\left|\varphi\right\rangle =\alpha\left|0\right\rangle +\beta\left|1\right\rangle$ by distributing it
throughout a $n$-party network. This is simply accomplished by
performing a Bell measurement between the initiator control qubit and the arbitrary
state qubit, such that the final state reads
\begin{equation}
\left|\psi\right\rangle =\frac{1}{\sqrt{2}}\left(\alpha\left|0_{L}\right\rangle +\beta\left|1_{L}\right\rangle \right).\label{eq:encoding2}
\end{equation}
This delocalization process offers a natural protection of the quantum information
under errors or losses of individual parties. In the same direction,
this construction can be useful for some settings in the context of quantum secret sharing \cite{Markham08,Hillery99}, where
a dealer (the control system) aims to share a secret between all the
constituents. The secret is defined by parameters $\alpha,\beta$
and only authorized sets of parties can, collaboratively, access to
it.


\subsection{Other applications}

We briefly discuss in this section some ideas for possible further applications of adding control to different tasks in a coherent way, although we do not discuss them in detail. We hope that this work motivates additional analysis and extensions for these and other applications.

\subsubsection{Superposition of QEC and QKD protocols}

Similarly to Sec. \ref{sec:encod}, one can generate superpositions of different encodings or codes such as e.g. quantum error correction (QEC) codes \cite{Dr2007,Lidar2009}. Consider two codes, $C_{1}$ and $C_{2}$, where e.g. one is particularly adequate for  correcting amplitude-bit errors and the other for correcting phase-bit errors. We conjecture that a superposition of both, i.e.
\begin{equation}
\left|\Psi\right\rangle=\alpha|\psi_{L}^{\left(C_{1}\right)}\rangle+\beta|\psi_{L}^{\left(C_{2}\right)}\rangle,
\end{equation}
might lead to advantages for correcting certain kind of errors under specific circumstances.

On the other hand, one can also consider Quantum Key Distribution (QKD) protocols \cite{Ekert91,Bennett2014}. The simplest case, the so-called BB84 protocol \cite{Bennett2014}, basically consists in Alice preparing a bit string of qubits in the $\left|0\right\rangle$ or $\left|1\right\rangle$ states out of two possible bases (e.g. $\sigma_{z}$ and $\sigma_{x}$ bases). The information of the bases is encoded in another bit string. The qubit bit string is sent to Bob, who randomly chooses a basis to measure each qubit. By classical publication of the chosen bases of each one, they can, by comparison, generate secure quantum keys protected from attacks. With our mechanisms, one can for instance, generate superposition of different BB84 protocols, in such a way that each constituent of the superposition involves a different BB84 with a different pair of bases, such as e.g. $\{\sigma_{z},\sigma_{x}\}$, $\{\sigma_{z},\sigma_{y}\}$ and $\{\sigma_{x},\sigma_{y}\}$. This might improve the protocol performance, including the number of qubits required or the stability under  errors.

\subsubsection{Sending of qubits}
We have considered in this paper sending of quantum information via teleportation. Following Sec. \ref{sec:csend}, we are able to generate states in superposition between teleporting some desired state and not teleporting any useful information (or not teleporting at all). This framework can be extended to directly sending ---or not--- a qubit via quantum channel. More precisely, the information carrier should be sent or not. When using ions or atoms as information carriers, this implies that the classical process to send the atom or not needs to be controlled - or alternatively an atom is always transmitted, either the one carrying information or some auxiliary one prepared in a dummy state. The latter solution however requires the usage of the channel also in cases where no quantum information is transmitted.

In the more realistsic setting where photons are used as information carriers, one should be able to add control at a quantum level. Following some of the existing techniques in cavity-QED with ions in cavities (see e.g. \cite{Northup2014}), where the internal state of the ion is mapped to the cavity field that eventually leaks out and is transmitted as a photon, one should be able to generate such a superposition state by just adding control to the mapping process via a second ion stored in the cavity. No photon will be generated (and hence transmitted) if the control qubit is in state $|0\rangle$.

\subsubsection{Distributed quantum sensing}

Quantum metrology \cite{Giovannetti2011,sekatski2019optimal} allows to carry out high precision measurements of certain physical quantities with an improved precision as compared to classical techniques. In particular, distributed quantum sensing \cite{sekatski2019optimal} consists of distributed multi-partite entangled sensor states, located and measured at different positions, in order to determine certain non-local physical quantities such as field gradients or higher moments of a scalar field. The quantum systems are prepared in particular states and evolve for a certain time, before they are finally measured in a certain way. With the techniques shown in this paper, one would be in principle able to generate superpositions at any step of the sensing process, including superposition between different experiments, e.g. between the measurement of some component of a constant field and the gradient on another component. We conjecture that the generation of these superpositions could lead to better efficiency and performance results for distributed quantum sensing procedures. Note however, that further extensions or modifications of our methods would be needed. While in principle we can establish coherent superpositions of different experiments, including the preparation of different states and running experiments for different times, at the level of measurements a problem arises. By definition, quantum metrology deals with the measurement of unknown states, as the desired (unknown) information of the quanity of interest is encoded in the states and revealed by measurements. The methods to add coherent control to measurements we presented in this work however require at least partial knowledge of the state to be measured, and are hence not directly applicable. 

\section{Summary and conclusions}\label{sec:conclusions}

In this paper we have shown how to provide quantum networks with a truly quantum functionality. This novel functionality allows network devices to handle or operate with coherent superpositions of different tasks.
The preparation of these superpositions is achieved by effectively controlling classical tasks in a coherent way. Adding explicit control to classical task is, in general, an impossible process. However, we have presented mechanisms, based on the control of quantum unitary operations, that can mimick the desired behavior. The crucial element is that the classical part is not controlled, but always performed, either acting usefully on the desired state or in vain on some dummy state. This mechanism involves the application of controlled-swap operations followed by a measurement of some auxiliary particles. Different tools arise from this approach, such as the possibility of effectively performing controlled-measurements, as well as e.g. the generation of superposition of sending or not sending information, or merging or not merging two graph states. Based on these tools, superposition of different tasks can be generated, either with or without external control. For this later case, we show procedures to suitably detach the control register while keeping the coherence and orthogonality of the state constituents.

Finally, we have shown different promising applications that emerge from our approach. Among them, one can highlight the possibility of preparing superpositions of states shared between different devices, superpositions of distributing quantum information through different paths or to different destinations, and superpositions of encoding information among different devices. These examples demonstrate possible advantages, most notable built-in robustness and favourable entanglement features. We hope that this work motivates future investigations of further possible extensions or applications of the truly quantum functionality of quantum networks introduced here.

\section*{Acknowledgements}
This work was supported by the Austrian Science Fund (FWF) through project P30937-N27. We are grateful to Vedran Dunjko for enlighting discussions on controlled measurements.

\bibliographystyle{apsrev4-1}
\bibliography{quantum_adressing}

\newpage
\onecolumngrid

\appendix
\numberwithin{equation}{section}

\section{Impossibility of coherently controlled measurements} \label{Apendcmeas}

We provide indications that a controlled known measurement performed in a coherent way on some unknown state is not realizable. Given a control and a target qubit, the desired effect of the operation is to obtain a coherent superposition between the target state being measured ---in case the control qubit is in $\left|1\right\rangle$--- and not measured ---when control qubit is in $\left|0\right\rangle$---, as explained in Sec. \ref{sec:cmeasur}.
For that purpose, consider two qubit registers. The first one, conceived as control register, is given in the state:
\begin{equation}
\left|\varphi\right\rangle _{c}=\alpha_{c}\left|0\right\rangle +\beta_{c}\left|1\right\rangle,
\end{equation}
where coefficients $\alpha_{c}$ and $\beta_{c}$ define the weights of the desired superposition. The measurement is performed in a controlled way on some target state $|\psi\rangle_{t}$. The desired effect for an arbitrary projective measurement $M=\{A_k\}$ is
\begin{equation}
|\varphi\rangle_c|\psi\rangle_t \rightarrow \{p_k, \alpha_0|0\rangle_c|\psi\rangle_t + \alpha_1|1\rangle|\psi_k\rangle\},\label{eq:measapen}
\end{equation}
with $\sum_k A_k^\dagger A_k =1$. One sees
that different branches arise from the measurement, each one corresponding
with each measurement outcome, with probability $p_{k}$.

\subsubsection{Observation 1}

In this first observation, we try to show that the required controlled measurement is not a valid quantum operation, by considering a general initial setting, and the desired final state. We see that, apparently, linearity of the operation is not guaranteed.

Consider again two qubit registers prepared in arbitrary states, $\left|\varphi\right\rangle _{c}=\alpha_{c}\left|0\right\rangle +\beta_{c}\left|1\right\rangle $
and $\left|\varphi\right\rangle _{t}=\alpha_{t}\left|0\right\rangle +\beta_{t}\left|1\right\rangle $
respectively, with $\left|\alpha_{i}\right|^{2}+\left|\beta_{i}\right|^{2}=1$.
The first qubit is conceived as the control register and the second
as the target. One aims to apply a controlled-measurement operation
between the two qubits. The desired effect of the controlled measurement
is given by Eq. \ref{eq:measapen}.
Consider the initial state
\begin{equation}
\rho_{0}=\left|\psi_{0}\right\rangle \left\langle \psi_{0}\right|,\label{eq:initialstate}
\end{equation}
where $\left|\psi_{0}\right\rangle $ is the composite system

\begin{align}
\left|\psi_{0}\right\rangle =\left|\varphi\right\rangle _{c}\left|\varphi\right\rangle _{t}=
\alpha_{c}\alpha_{t}\left|00\right\rangle +\alpha_{c}\beta_{t}\left|01\right\rangle +\beta_{c}\alpha_{t}\left|10\right\rangle +\beta_{c}\beta_{t}\left|11\right\rangle =
a\left|00\right\rangle +b\left|01\right\rangle +c\left|10\right\rangle +d\left|11\right\rangle,
\end{align}
where coefficients have been relabelled for simplicity. We assume a projective
measurement performed in the computational basis. Since two different branches arise from the measurement (see Eq. \ref{eq:measapen}), the desired final state has to be described as a classical mixture of two states, each one corresponding to each of the
possible outcomes of the projective measurement:

\begin{equation}
\rho_{f}=p_{1}\left|\psi_{1}\right\rangle \left\langle \psi_{1}\right|\otimes\left|1\right\rangle _{m}\left\langle 1\right|+p_{2}\left|\psi_{2}\right\rangle \left\langle \psi_{2}\right|\otimes\left|2\right\rangle _{m}\left\langle 2\right|,\label{eq:finalstate}
\end{equation}
where register $m$ defines the apparatus state, i.e. the outcome of the measurement. Besides,  $p_{1}+p_{2}=1$, where 

\begin{equation}
\left|\psi_{1}\right\rangle =a\left|00\right\rangle +b\left|01\right\rangle +\sqrt{cc^{*}+dd^{*}}\left|10\right\rangle=\alpha_{c}\left|0\right\rangle_{c}\left|\varphi\right\rangle _{t}+ \left|\beta_{c}\right|\left|1\right\rangle_{c}\left|0\right\rangle_t,
\end{equation}

\begin{equation}
\left|\psi_{2}\right\rangle =a\left|00\right\rangle +b\left|01\right\rangle +\sqrt{cc^{*}+dd^{*}}\left|11\right\rangle=\alpha_{c}\left|0\right\rangle_{c}\left|\varphi\right\rangle _{t}+ \left|\beta_{c}\right|\left|1\right\rangle_{c}\left|1\right\rangle_t,
\end{equation}
with $p_{1}=\frac{cc^{*}}{cc^{*}+dd^{*}}=\left|\alpha_{t}\right|^{2}$   and   $p_{2}=\frac{dd^{*}}{cc^{*}+dd^{*}}=\left|\beta_{t}\right|^{2}$.

Observe that the desired superposition is found for each term
of the mixture. For each individual term, the measurement is performed if the control qubit is in the state $\left|1\right\rangle $ with probability
$\left|\beta_{c}\right|^{2}$, where $\beta_{c}=\sqrt{\left|c\right|^{2}+\left|d\right|^{2}}$.
The target qubit remains untouched if the control
state is $\left|0\right\rangle $ (with probability $\left|\alpha_{c}\right|^{2}$).
Therefore, in each branch of the measurement,  a coherent superposition
of measuring and not measuring the target qubit (with their corresponding
probabilities) is found. It is however straightforward to see that
the map from Eq. (\ref{eq:initialstate}) to Eq. (\ref{eq:finalstate}), i.e. $\xi\left(\rho_{0}\right)=\rho_{f}$, is not a linear transformation,
i.e. $\xi\left(\sum_{i}p_{i}\rho_{i}\right)\neq\sum_{i}p_{i}\xi(\rho_{i})$.
By direct inspection one observes that the off-diagonal terms of Eq.~(\ref{eq:finalstate})
related with the control qubit being in the state $\left|1\right\rangle$
are not linear functions of the elements of Eq.~(\ref{eq:initialstate}).
This might indicate that the map is not linear and, consequently, it
is not a valid quantum operation.

\subsubsection{Observation 2}

In this second observation, we assume that the controlled measurement is linear and we find inconsistencies that might indicate that the linearity assumption is not correct.

First, consider the desired effect of a controlled measurement on
different states. For simplicity, we choose the control state to be
in the $\left|\psi\right\rangle _{c}=\left|+\right\rangle $ state,
i.e. same weights of the desired superposition. Besides, we consider
a projective measurement in the computational basis. The initial state
reads
\begin{equation}
\left|\varphi\right\rangle =\left|+\right\rangle _{c}\left|\psi\right\rangle _{t}=\frac{1}{\sqrt{2}}\left(\left|0\right\rangle _{c}\left|\psi\right\rangle _{t}+\left|1\right\rangle _{c}\left|\psi\right\rangle _{t}\right).
\end{equation}
The desired effect of the operation for different target states and
the different measurement branches is:
\begin{itemize}
\item $\left|\psi\right\rangle _{t}=\left|0\right\rangle $
\end{itemize}
\begin{equation}
\left|\varPhi_{00}\right\rangle =\frac{1}{\sqrt{2}}\left(\left|0\right\rangle _{c}\left|0\right\rangle _{t}+\left|1\right\rangle _{c}\left|0\right\rangle _{t}\right)\,\,\,\text{with}\,\,p=1
\end{equation}

\begin{equation}
\left|\varPhi_{01}\right\rangle =\frac{1}{\sqrt{2}}\left(\left|0\right\rangle _{c}\left|0\right\rangle _{t}+\left|1\right\rangle _{c}\left|1\right\rangle _{t}\right)\,\,\,\text{with}\,\,p=0
\end{equation}

\begin{itemize}
\item $\left|\psi\right\rangle _{t}=\left|1\right\rangle $
\end{itemize}
\begin{equation}
\left|\varPhi_{10}\right\rangle =\frac{1}{\sqrt{2}}\left(\left|0\right\rangle _{c}\left|1\right\rangle _{t}+\left|1\right\rangle _{c}\left|0\right\rangle _{t}\right)\,\,\,\text{with}\,\,p=0
\end{equation}

\begin{equation}
\left|\varPhi_{11}\right\rangle =\frac{1}{\sqrt{2}}\left(\left|0\right\rangle _{c}\left|1\right\rangle _{t}+\left|1\right\rangle _{c}\left|1\right\rangle _{t}\right)\,\,\,\text{with}\,\,p=1
\end{equation}

\begin{itemize}
\item $\left|\psi\right\rangle _{t}=\left|+\right\rangle $
\end{itemize}
\begin{equation}
\left|\varPhi_{+0}\right\rangle =\frac{1}{\sqrt{2}}\left(\left|0\right\rangle _{c}\left|+\right\rangle _{t}+\left|1\right\rangle _{c}\left|0\right\rangle _{t}\right)\,\,\,\text{with}\,\,p=\frac{1}{2}
\end{equation}

\begin{equation}
\left|\varPhi_{+1}\right\rangle =\frac{1}{\sqrt{2}}\left(\left|0\right\rangle _{c}\left|+\right\rangle _{t}+\left|1\right\rangle _{c}\left|1\right\rangle _{t}\right)\,\,\,\text{with}\,\,p=\frac{1}{2}
\end{equation}

\begin{itemize}
\item $\left|\psi\right\rangle _{t}=\left|-\right\rangle $
\end{itemize}
\begin{equation}
\left|\varPhi_{-0}\right\rangle =\frac{1}{\sqrt{2}}\left(\left|0\right\rangle _{c}\left|-\right\rangle _{t}+\left|1\right\rangle _{c}\left|0\right\rangle _{t}\right)\,\,\,\text{with}\,\,p=\frac{1}{2}
\end{equation}

\begin{equation}
\left|\varPhi_{-1}\right\rangle =\frac{1}{\sqrt{2}}\left(\left|0\right\rangle _{c}\left|-\right\rangle _{t}-\left|1\right\rangle _{c}\left|1\right\rangle _{t}\right)\,\,\,\text{with}\,\,p=\frac{1}{2}
\end{equation}
Consider now the state $\rho_{t}=\frac{1}{2}\mathbb{1}$. The
initial state now reads $\rho_{0}=\left|+\right\rangle _{c}\left\langle +\right|\otimes\frac{1}{2}\mathbb{1}_{t}$.
We analyze two different cases arising from this:
\begin{enumerate}
\item Decompose $\rho_{t}=\frac{1}{2}\mathbb{1}=\frac{1}{2}\left|0\right\rangle \left\langle 0\right|+\frac{1}{2}\left|1\right\rangle \left\langle 1\right|$.
\end{enumerate}
From this decomposition and from the initial state $\rho_{0} $, assuming that the controlled-measurement we consider is linear, the operation applied to $\rho_{0}$ leads to
\begin{equation}
\rho =\frac{1}{2}\left|\varPhi_{00}\right\rangle \left\langle \varPhi_{00}\right|\otimes\left|0\right\rangle _{M}\left\langle 0\right|+\frac{1}{2}\left|\varPhi_{11}\right\rangle \left\langle \varPhi_{11}\right|\otimes\left|1\right\rangle _{M}\left\langle 1\right|, \label{eq:cmeas11}
\end{equation}
where the states $\left|\varPhi _{ij}\right\rangle$ are the desired
actions previously defined, and the register $M$ defines the apparatus
state, i.e. the measurement outcome.
\begin{enumerate}
\item Decompose $\rho_{t}=\frac{1}{2}\mathbb{1}=\frac{1}{2}\left|+\right\rangle \left\langle +\right|+\frac{1}{2}\left|-\right\rangle \left\langle -\right|$.
\end{enumerate}
Analogously, for this decomposition and the effect on different target states shown above, the final state reads

\begin{equation}
\rho =\frac{1}{4}\left|\varPhi_{+0}\right\rangle \left\langle \varPhi_{+0}\right|\otimes\left|0\right\rangle _{M}\left\langle 0\right|+\frac{1}{4}\left|\varPhi_{+1}\right\rangle \left\langle \varPhi_{+1}\right|\otimes\left|1\right\rangle _{M}\left\langle 1\right|+\frac{1}{4}\left|\varPhi_{-0}\right\rangle \left\langle \varPhi_{-0}\right|\otimes\left|0\right\rangle _{M}\left\langle 0\right|+\frac{1}{4}\left|\varPhi_{-1}\right\rangle \left\langle \varPhi_{-1}\right|\otimes\left|1\right\rangle _{M}\left\langle 1\right|. \label{eq:cmeas22}
\end{equation}
One can easily see that expressions Eq. (\ref{eq:cmeas11}) and Eq. (\ref{eq:cmeas22}) do not match. This contradiction might directly indicate that our assumption is wrong, and the controlled-measurement is indeed not linear. 

\section{Controlled sending} \label{Apendcsend}

In this appendix, we provide the formalization of the controlled-sending introduced in 
Sec.~\ref{sec:csend} by means of quantum teleportation. Crucially, we observe that the coefficients of the superposition do not change throughout the process. The initial state reads
\begin{align}
\left|\Psi\right\rangle _{in}=\left(\alpha_{0}\left|0\right\rangle _{c}+\alpha_{1}\left|1\right\rangle _{c}\right)\left|\psi\right\rangle _{a_{1}}\left|\Phi^{+}\right\rangle _{a_{2}b}\left|0\right\rangle _{ax_{1}}\left|+\right\rangle _{ax_{2}},
\end{align}
where party $A$ possesses five qubits, the control register $c$,
one qubit in the arbitrary state $\left|\psi\right\rangle _{a_{1}}=\alpha\left|0\right\rangle +\beta\left|1\right\rangle $
(to be sent), another one $a_{2}$ sharing a Bell state with party
$B$, and two auxiliary qubits ($ax_{1}$ and $ax_{2}$) initialized
in the states $\left|0\right\rangle _{ax_{1}}$ and $\left|+\right\rangle _{ax_{2}}$
respectively. Party $B$ just possesses the qubit $b$ of the Bell
state $\left|\Phi^{+}\right\rangle _{a_{2}b}=\frac{1}{\sqrt{2}}\left(\left|00\right\rangle _{a_{2}b}+\left|11\right\rangle _{a_{2}b}\right)$.
First, the controlled-swap operation of Eq. (\ref{eq:cswap}) is applied simultaneously between qubits $a_{1}$ and $ax_{1}$, and qubits $a_{2}$ and $ax_{2}$, such that
the global states evolves to 
\begin{align}
\left|\Psi\right\rangle =\alpha_{0}\left|0\right\rangle _{c}\left|0\right\rangle _{ax_{1}}\left|+\right\rangle _{ax_{2}}\left|\Phi^{+}\right\rangle _{a_{2}b}\left|\psi\right\rangle _{a_{1}} +  
\frac{1}{\sqrt{2}}\alpha_{1}\left|1\right\rangle _{c}\left(\alpha\left(\left|000\right\rangle _{ax_{1}ax_{2}b}+ 
\left|011\right\rangle _{ax_{1}ax_{2}b}\right)+\beta\left(\left|100\right\rangle _{ax_{1}ax_{2}b}+\left|111\right\rangle _{ax_{1}ax_{2}b}\right)\right)\left|0\right\rangle _{a_{1}}\left|+\right\rangle _{a_{2}}.
\end{align}
Second, a Bell measurement is always applied by party $A$ between
systems $ax_{1}$ and $ax_{2}$. Assuming the outcome of this measurement
is $\Phi_i$,
and taking into account the decomposition
\begin{align}
\left|00\right\rangle =\frac{1}{\sqrt{2}}\left(\left|\Phi^{+}\right\rangle +\left|\Phi^{-}\right\rangle \right)\,,\,\left|11\right\rangle =\frac{1}{\sqrt{2}}\left(\left|\Phi^{+}\right\rangle -\left|\Phi^{-}\right\rangle \right),
\end{align}
the state evolves to
\begin{align}
\left|\Psi\right\rangle =\alpha_{0}\left|0\right\rangle _{c}\frac{1}{2}\left|\Phi_i\right\rangle _{ax_{1}ax_{2}}\left|\Phi^{+}\right\rangle _{a_{2}b}\left|\psi\right\rangle _{a_{1}}+\alpha_{1}\left|1\right\rangle _{c}\frac{1}{\sqrt{2}}\left(\frac{1}{\sqrt{2}}\alpha\left|\Phi_i\right\rangle _{ax_{1}ax_{2}}\left|0\right\rangle _{b}+\frac{1}{\sqrt{2}}\beta\left|\Phi_i\right\rangle _{ax_{1}ax_{2}}\left|1\right\rangle _{b}\right)\left|0\right\rangle _{a_{1}}\left|+\right\rangle _{a_{2}}.
\end{align}
In case another measurement outcome is found, a  controlled unitary correction operation on qubit $b$ of the form $\left|0\right\rangle_{c} \left\langle 0\right|\otimes\mathbb{1}+\left|1\right\rangle_{c} \left\langle 1\right|\otimes\sigma_{i}^{b}$ is necessary. Note that, since the control register belongs to party $A$, this correction can always be introduced by party $B$ applying locally the unitary $\sigma_{i}$ on qubit $b$, followed by party $A$ applying the controlled unitary $\left|0\right\rangle_{c} \left\langle 0\right|\otimes\sigma_{i}^\intercal+\left|1\right\rangle_{c} \left\langle 1\right|\otimes\mathbb{1}$ on qubit $a_{2}$. Finally, after re-normalization  the desired state is found:

\begin{equation}
\left|\Psi\right\rangle _{f}=\left(\alpha_{0}\left|0\right\rangle _{c} \left|\psi\right\rangle _{a_{1}} \left| \Phi^+\right\rangle _{a_{2} b}+\alpha_{1}\left|1\right\rangle _{c} \left|0\right\rangle _{a_{1}}\left|+\right\rangle _{a_{2}} \sigma_i\left|\psi\right\rangle _{b} \right) 
\left|\Phi_i\right\rangle _{ax_{1}ax_{2}}.\label{eq:sendingfinalaux}
\end{equation}
Therefore, the weights of the superposition remain unchanged and the
desired target state is generated, where the state $\left|\psi\right\rangle$ is teleported in case of the control qubit is in $\left|1\right\rangle$, and the state $\left|\psi\right\rangle$ and the shared Bell pair are kept untouched otherwise.

Additionally, if one is not interested on keeping the initial Bell state between parties $A$ and $B$, qubit $a_{2}$ can be measured in the computational basis. Assuming the outcome $\left|0\right\rangle $ is obtained, the resulting
state is
\begin{align}
\left|\Psi\right\rangle _{f}=\left(\alpha_{0}\left|0\right\rangle _{c}\left|0\right\rangle _{b}\left|\psi\right\rangle _{a_{1}}+\alpha_{1}\left|1\right\rangle _{c}\left|\psi\right\rangle _{b}\left|0\right\rangle _{a_{1}}\right)\left|\Phi^{+}\right\rangle _{ax_{1}ax_{2}}\left|0\right\rangle _{a_{2}},
\end{align}
where the state has been re-normalized. 

\section{Controlled merging} \label{Apendcmerg}
In this appendix, we provide a detailed mathematical description of
the controlled-merging of graph states of Sec. \ref{sec:cmerg}. The initial state consists
of two graph states, an auxiliary qubit, and a control register
(see Fig. \ref{fig:cmerg}) 

\begin{equation}
\left|\Psi\right\rangle _{in}=\alpha_{0}\left|0\right\rangle _{c}\left|G_{1}\right\rangle \left|G_{2}\right\rangle \left|+\right\rangle _{aux}+\alpha_{1}\left|1\right\rangle _{c}\left|G_{1}\right\rangle \left|G_{2}\right\rangle \left|+\right\rangle _{aux},\label{eq:apmerg1}
\end{equation}
where $\left|G_{i}\right\rangle =\frac{1}{\sqrt{2}}\left(\left|0\right\rangle _{a_{i}}\left|G_{i}/a_{i}\right\rangle +\left|1\right\rangle _{a_{i}}\prod\sigma_{z}^{N_{a_{i}}}\left|G_{i}/a_{i}\right\rangle \right)$.
Qubits $a_{1},a_{2}$ and $aux$ belong to the same device $A$,
which also possesses the control register. We analyze both branches
of Eq. (\ref{eq:apmerg1}) independently. First, consider the case
where the control register is in $\left|0\right\rangle _{c}$ the
state. In this case, a swap operation is applied between qubits
$a_{2}$ and $aux$. Note that, in the rest of the paper, we consider
that the swap is performed if the control register is in $\left|1\right\rangle _{c}$.
We change here this criteria for convenience. After the swap operation
is applied, the state of the $\left|0\right\rangle _{c}$ branch reads:
\[
\frac{1}{\sqrt{2}}\alpha_{0}\left|0\right\rangle _{c}\left|G_{1}\right\rangle \left(\left|+\right\rangle _{a_{2}}\left|G_{2}/aux\right\rangle \left|0\right\rangle _{aux}+\left|+\right\rangle _{a_{2}}\prod\sigma_{z}^{N_{a_{aux}}}\left|G_{2}/aux\right\rangle \left|1\right\rangle _{aux}\right).
\]
Finally, the merging measurement $\left\{ P_{0}=\left|0\right\rangle \left\langle 00\right|+\left|1\right\rangle \left\langle 11\right|,P_{1}=\left|0\right\rangle \left\langle 01\right|+\left|1\right\rangle \left\langle 10\right|\right\} $
is performed between qubits $a_{1}$ and $a_{2}$. By expanding
$\left|G_{1}\right\rangle $, one finds, for both measurement outcomes: 

\begin{equation}
\frac{1}{2\sqrt{2}}\alpha_{0}\left|0\right\rangle _{c}\left(\left|0\right\rangle _{\tilde{a_{1}}}\left|G_{1}/\tilde{a_{1}}\right\rangle +\left|1\right\rangle _{\tilde{a_{1}}}\prod\sigma_{z}^{N_{\tilde{a_{1}}}}\left|G_{1}/\tilde{a_{1}}\right\rangle \right)\left(\left|0\right\rangle _{aux}\left|G_{2}/aux\right\rangle +\left|1\right\rangle _{aux}\prod\sigma_{z}^{N_{a_{aux}}}\left|G_{2}/aux\right\rangle \right),\label{eq:apmerg2}
\end{equation}
where $\tilde{a_{1}}$ is the resulting qubit of merging $a_{1}$
and $a_{2}$.

On the other hand, for the branch of Eq. (\ref{eq:apmerg1}) corresponding
to $\left|1\right\rangle _{c}$, the swap operation is not applied.Therefore,
the next step is the merging operation. The state before
the measurement is 
\[
\frac{1}{2}\alpha_{1}\left|1\right\rangle _{c}\left(\left|0\right\rangle _{a_{1}}\left|G_{1}/a_{1}\right\rangle +\left|1\right\rangle _{a_{1}}\prod\sigma_{z}^{N_{a_{1}}}\left|G_{1}/a_{1}\right\rangle \right)\left(\left|0\right\rangle _{a_{2}}\left|G_{2}/a_{2}\right\rangle +\left|1\right\rangle _{a_{2}}\prod\sigma_{z}^{N_{a_{2}}}\left|G_{2}/a_{2}\right\rangle \right)\left|+\right\rangle _{aux}.
\]
After merging qubits $a_{1}$ and $a_{2}$ into $\tilde{a_{1}}$,
the resulting state for the outcome corresponding to $P_{0}$ is 

\begin{equation}
\frac{1}{2}\alpha_{1}\left|1\right\rangle _{c}\left(\left|0\right\rangle _{\tilde{a_{1}}}\left|G_{1}\cup G_{2}/\tilde{a_{1}}\right\rangle +\left|1\right\rangle _{\tilde{a_{1}}}\prod\sigma_{z}^{N_{\tilde{a_{1}}}}\left|G_{1}\cup G_{2}/\tilde{a_{1}}\right\rangle \right)\left|+\right\rangle _{aux},\label{eq:apmerg3}
\end{equation}
where $N_{\tilde{a_{1}}}=N_{a_{1}}\cup N_{a_{2}}.$ In case the measurement
outcome corresponds to $P_{1}$, the state is 

\[
\frac{1}{2}\alpha_{1}\left|1\right\rangle _{c}\left(\prod\sigma_{z}^{N_{a_{2}}}\left|0\right\rangle _{\tilde{a_{1}}}\left|G_{1}\cup G_{2}/\tilde{a_{1}}\right\rangle +\left|1\right\rangle _{\tilde{a_{1}}}\prod\sigma_{z}^{N_{a_{1}}}\left|G_{1}\cup G_{2}/\tilde{a_{1}}\right\rangle \right)\left|+\right\rangle _{aux}.
\]
Therefore, in case one obtains the merging measurement outcome corresponding
to $P_{1}$, a controlled correction operation of the form $\left|0\right\rangle_{c} \left\langle 0\right|\otimes\mathbb{1}+\left|1\right\rangle_{c} \left\langle 1\right|\otimes\prod\sigma_{z}^{N_{a_{2}}}$ is required
to recover the state of Eq. (\ref{eq:apmerg3}). Note that the controlled correction is intended to have no influence on the other branch of the superposition, such that the probabilities of pretending to measure each branch individually are such that the weights in the superposition do not change. This is possible due to the orthogonality of the constituents of the superposition.

Taking into account Eq. (\ref{eq:apmerg2}) and Eq. (\ref{eq:apmerg3}),
and by simply relabelling qubits $aux\rightarrow a_{2}$ and $\tilde{a_{1}}\rightarrow a_{1}$,
the final global state after re-normalization reads

\begin{equation}
\left|\Psi\right\rangle _{f}=\alpha_{0}\left|0\right\rangle _{c}\left|G_{1}\right\rangle \left|G_{2}\right\rangle +\alpha_{1}\left|1\right\rangle _{c}\left|G_{1}\cup G_{2}\right\rangle \left|+\right\rangle _{a_{2}},\label{eq:cmerg4}
\end{equation}
where a superposition between two graph states unaltered and two graph
states merged is found. Observe that, crucially, superposition weights remain unchanged
through the process. Note also that qubit $a_{2}$ of the second
branch is now some auxiliary qubit that has been relabelled for
convenience. 

\end{document}